\documentclass[journal]{IEEEtran}
\IEEEoverridecommandlockouts
\usepackage{booktabs}
\usepackage{multirow}
\usepackage{cite}
\usepackage{amsmath,amssymb,amsfonts}
\usepackage{algorithmic}
\usepackage{graphicx}
\usepackage{textcomp}
\usepackage{balance}
\usepackage[switch]{lineno}
\usepackage{array}
\usepackage{microtype}
\microtypecontext{expansion=basictext,spacing=nonfrench}
\UseMicrotypeSet[protrusion]{basicmath} 

\usepackage{mathtools, amsmath, amssymb}

\usepackage{hyperref}
\hypersetup{
    colorlinks=true,
    linkcolor=red,
    filecolor=magenta,      
    urlcolor=blue,
    citecolor=blue
}

\usepackage{xcolor}
\def\BibTeX{{\rm B\kern-.05em{\sc i\kern-.025em b}\kern-.08em
    T\kern-.1667em\lower.7ex\hbox{E}\kern-.125emX}}
\begin{document}
\title{Development of Domain-Invariant Visual Enhancement and Restoration (DIVER) Approach  for Underwater Images}
\author{Rajini Makam $^{1*}$, Sharanya Patil$^{2}$, Dhatri Shankari T M$^{2}$, Suresh Sundaram$^{1}$, Narasimhan Sundararajan$^{3}$

    \thanks{$^{1}$Rajini Makam and Suresh Sundaram are with the Department of Aerospace Engineering, Indian Institute of Science, Bangalore, India.
    {\tt\small \{rajinimakam@iisc.ac.in, vssuresh@iisc.ac.in\}}}%
    \thanks{$^{2}$ Sharanya Patil and Dhatri Shankari T M were interns at Department of Aerospace, Indian Institute of Science, Bangalore, India. {\tt\small \{sharanyapatil23@gmail.com, dhatristm2003@gmail.com\}}}%
     \thanks{$^{3}$ Retired Professor, Nanyang Technological University, Singapore. 
    {\tt\small \{ensundara@ntu.edu.sg\}}}%
  }  

\maketitle

\begin{abstract}
Underwater imaging suffers from severe degradation due to wavelength-dependent attenuation, scattering, and illumination non-uniformity, which vary significantly across water types and depths. In this paper, we propose an unsupervised Domain-Invariant Visual Enhancement and Restoration (DIVER) architecture that combines empirical corrections and physics-guided modeling for robust underwater image enhancement. First, DIVER uses either IlluminateNet for adaptive luminance enhancement or the Spectral Equalization Filter for spectral normalization. Subsequently, the Adaptive Optical Correction Module refines hue and contrast through channel-adaptive filtering, while Hydro-OpticNet employs a physics-constrained learning approach to compensate for backscatter and wavelength-dependent attenuation. The tunable parameters in IlluminateNet and Hydro-OpticNet are updated using unsupervised learning with a composite loss function. DIVER is evaluated on eight diverse datasets encompassing shallow, deep, and highly turbid environments, including both naturally low-lit and artificially illuminated scenarios, using reference and non-reference metrics. While SOTA methods such as WaterNet, UDNet, and Phaseformer show reasonable performance in shallow-water scenarios, they degrade significantly in deep, unevenly illuminated, or artificially lit environments. In contrast, DIVER consistently achieves best or near-best performance across all datasets, demonstrating domain-invariant capability. DIVER achieves at least 9\% improvement over SOTA in terms of UCIQE. On the low-light SeaThru dataset, where color-palette references enable direct evaluation of color restoration accuracy, DIVER shows at least a 4.9\% reduction in GPMAE compared to existing SOTA methods. Beyond visual quality metrics, DIVER also improves robotic perception, enhancing keypoint repeatability and matching performance using ORB descriptors. These results show that DIVER is robust and domain-invariant across diverse underwater environments.  \href{https://github.com/AIRLabIISc/DIVER}{https://github.com/AIRLabIISc/DIVER}.

\end{abstract}


\IEEEpeerreviewmaketitle

\section{Introduction}

Underwater imaging plays a pivotal role across diverse marine and industrial applications, including ecological monitoring, habitat mapping, archaeological exploration, defense operations, and inspection of submerged infrastructure. High-quality underwater images enable precise visual perception for tasks such as coral reef health assessment, seabed characterization, and object detection by Autonomous Underwater Vehicles (AUVs) and Remotely Operated Vehicles (ROVs), which rely on robust vision systems for navigation and obstacle avoidance \cite{Yang2019review}, \cite{dawkins2024}. However, obtaining clear underwater imagery remains challenging due to wavelength-dependent light absorption, scattering by suspended particles, turbidity, and non-uniform illumination \cite{shafuda2025, akkaynak2019}. These phenomena collectively cause color distortion, reduced contrast, and structural detail loss whose effects differ fundamentally from terrestrial imaging.

Traditional enhancement techniques improve underwater images by manipulating pixel intensities or frequency components without explicitly modeling underwater light propagation. Contrast Limited Adaptive Histogram Equalization (CLAHE)~\cite{hitam2013} enhances local contrast and reveals finer details, but it often amplifies noise and can introduce unnatural brightness, especially in homogeneous or low-texture underwater regions . Several model-based approaches further incorporate handcrafted priors to approximate underwater degradation. IBLA~\cite{peng2017} estimates scene depth from image blurriness and light absorption, whereas DCP~\cite{he2010} and its underwater variant UDCP~\cite{drews2013} use transmission-map estimation adapted from atmospheric dehazing to suppress haze and correct color casts. More recent enhancements, such as MLLE~\cite{MMLE2022}, offer local color and contrast correction but remain heuristic and sensitive to water-type variations.  Although computationally efficient, these approaches remain heuristic and often yield over-enhancement, amplified noise, or inaccurate color correction under severe scattering, strong turbidity, or wavelength-selective attenuation. Their reliance on handcrafted assumptions and approximate depth cues further limits robustness across diverse underwater domains.

With the rise of deep learning, data-driven models have significantly advanced underwater image enhancement. Supervised CNN and GAN-based methods such as WaterNet~\cite{li2019} learn nonlinear mappings between degraded and reference images, while more recent hybrid architectures like P2CNet~\cite{rao2023} integrate physical priors to improve stability and color fidelity. Although these approaches deliver strong visual enhancement, they rely on large paired datasets, which are rarely available underwater.  To reduce reliance on paired data, semi-supervised methods such as Xiao et al.~\cite{Xiao2025} leverage depth-map consistency and unlabeled samples for training, but their performance remains sensitive to dataset-specific spectral statistics and illumination conditions. UDNet~\cite{saleh2025adaptive}, an unsupervised encoder–decoder architecture, enhances structural details and suppresses scattering artifacts by leveraging adaptive priors and reconstruction-based constraints. Despite being more flexible than supervised models, UDNet still struggles to recover heavily attenuated warm colors, particularly red and yellow channels in deep-sea conditions.

Aforementioned enhancement and restoration approaches struggle to generalize across heterogeneous underwater domains such as shallow waters, deep oceanic regions, and highly turbid environments. Their reliance on dataset-specific priors, fixed color assumptions, or domain-limited learning prevents consistent performance when appearance statistics shift across water types.  Further, a fundamental challenge in underwater image restoration is the absence of ground-truth clean images, since clear underwater scenes cannot be captured directly due to inherent medium distortions. This lack of paired supervision renders traditional supervised learning approaches unreliable and highly domain-specific, motivating the need for unsupervised and physics-guided learning frameworks capable of generalizing without reference labels.

In this paper, we propose an unsupervised Domain-Invariant Visual Enhancement and Restoration architecture, called DIVER. The proposed unsupervised learning addresses luminance loss, chromatic imbalance, scattering-induced haze, and wavelength-dependent attenuation across diverse water conditions. The key contributions are summarized as follows:
DIVER integrates empirical correction and physics-guided modeling to robustly restore underwater images across shallow tropical, deep oceanic, and highly turbid environments—without the need for ground-truth labels. A luminance assessment module directs each input either to IlluminateNet for low-light enhancement or to Spectral Equalization Filter (SEF) for well-lit scenes, ensuring consistent brightness correction under both natural and artificial illumination. The Adaptive Optical Correction Module (AOCM) enhances hue uniformity, local contrast, and structural clarity while suppressing chromatic speckle artifacts prevalent in sediment-rich waters. Hydro-OpticNet, consisting of VeilNet and AttenNet, learns depth-conditioned backscatter removal and wavelength-dependent attenuation compensation, producing radiometrically consistent and perceptually balanced scene radiance. DIVER is evaluated across eight datasets capturing low-lit, artificially illuminated, deep-water, and high-turbidity scenarios. It outperforms ten state-of-the-art enhancement methods including IBLA~\cite{peng2017}, DCP~\cite{he2010}, UDCP~\cite{drews2013}, ULAP~\cite{song2018}, P2CNet~\cite{rao2023}, Phaseformer~\cite{khan2025}, WaterNet~\cite{li2019}, UDNet~\cite{saleh2025adaptive}, Spectroformer, and the U-Shape Transformer~\cite{peng2023}. DIVER achieves consistently better performance across both reference-based metrics such as PSNR and SSIM (when ground-truth labels are available), and no-reference perceptual metrics including UCIQE, UIQM, BRISQUE, and GPMAE (when visual quality can be assessed). The framework delivers best or near-best scores across paired and unpaired datasets, demonstrating reliable enhancement effectiveness and robust generalization in diverse underwater conditions.  In particular, UCIQE and UIQM show clear gains across all datasets. UCIQE gains at least 22\% over traditional methods, 24\% over transformer-based models, and 34\% over UDNet. On a low-lit SeaThru dataset, GPMAE improves by at least 4.9\%, confirming more accurate color fidelity due to reliable angular-error reduction. Feature-matching experiments using ORB descriptors show improved keypoint repeatability and match accuracy, demonstrating that DIVER not only improves visual quality but also strengthens actionable perception needed for autonomous underwater navigation.

The paper is structured as follows: Section \ref{s2} provides an overview of the existing work in the field of underwater image enhancement. Section \ref{s3} presents the proposed DIVER architecture. Results obtained from the DIVER architecture are discussed in Section \ref{s4}. Section \ref{s5} provides the conclusion of the work.

\section{Related Work} \label{s2}

Underwater image enhancement has significantly advanced in recent years, especially with the introduction of new comprehensive datasets and novel algorithms. Due to the unique challenges posed by aquatic environments, such as wavelength-dependent light absorption and scattering, underwater image enhancement has attracted widespread attention. This section covers the evolution of underwater image enhancement techniques starting with traditional methods to more recent advancements where these approaches can be broadly classified into two major categories: traditional underwater image enhancement methods and deep learning for underwater image enhancement.

\subsection{Traditional Underwater Image Enhancement Methods}

Classical enhancement approaches broadly include hardware-based imaging techniques and software-based image processing algorithms. Early hardware solutions such as polarization-based imaging improved visibility but were impractical due to the need for multiple synchronized captures. Contrast stretching and histogram equalization, enhance visibility but often introduce artifacts because they do not account for underwater light absorption and scattering. Methods such as ULAP~\cite{song2018} introduce wavelength-dependent luminance adaptation, and RGHS~\cite{huang2018} enhances global contrast through guided histogram stretching.

Several traditional methods attempt improvement without explicitly modeling underwater optics. Histogram-based techniques such as CLAHE~\cite{hitam2013} boost local contrast but may amplify noise, whereas other methods like {L$^2$UWE} ~\cite{marques2020} utilize contrast information for low-lit image correction but produce overly bright regions. Retinex-based models~\cite{lin2023underwater} correct illumination inconsistencies but often struggle with accurate color fidelity. Fusion-based approaches combine multiple corrected versions of an image to improve contrast and texture~\cite{liang2025under}, yet they still fail under severe wavelength-dependent attenuation, particularly in dark or turbid waters. Overall, traditional methods degrade in challenging environments due to their reliance on handcrafted assumptions, limited handling of color attenuation, and inability to adapt across heterogeneous underwater domains.

\subsection{Deep Learning for Underwater Image Enhancement}

The performance of computer vision algorithms for underwater image classification and scene reconstruction is significantly hindered by strong color distortions, which adversely affect both deep neural networks and traditional feature-based methods. While classical techniques often fail to cope with the nonlinear and depth-dependent degradations inherent to underwater imagery, deep learning models have demonstrated improved robustness by learning color correction patterns from large-scale datasets~\cite{peng2022}. Encoder–decoder architectures such as UDNet~\cite{saleh2025adaptive} refine structural details through hierarchical feature learning, yet they frequently struggle to recover severely attenuated warm colors (reds and yellows) in deep or turbid waters. Physics-based enhancement networks further address this challenge by modeling light–water interactions, allowing explicit estimation of image formation parameters to reverse color shifts and suppress scattering-induced noise~\cite{chandrasekar2024}. For example, vignetting-correction–based enhancement schemes designed for AUV imagery explicitly account for artificial illumination geometry, improving radiometric uniformity \cite{Vi2025}. Other supervised approaches, such as the four-stage color decomposition method in ~\cite{cong2024underwater}, offer fine-grained color refinement but remain sensitive to illumination variations. Likewise, UIEB-trained WaterNet~\cite{li2019} produces visually appealing results but exhibits inconsistent color recovery across diverse water types due to dataset-specific biases.

Generative methods, including GAN-based approaches for paired and unpaired training-~\cite{islam2020}, such as WaterGAN~\cite{li2017}, attempt to learn domain mappings directly from data; however, they are data-intensive and often generalize poorly when deployed in real-world conditions with severe turbidity or extreme spectral shifts unmatched by the training domain. Wavelet-driven enhancement networks, such as those utilizing multiscale frequency decomposition~\cite{weidong2025} or hybrid wavelet-transformer architectures like MixRformer~\cite{li2025mixrformer}, improve color and detail restoration but can introduce fusion artifacts due to imperfect sub-band recombination. Recent supervised fusion frameworks, exemplified by the Turbidity Suppression Fusion (TSF) model~\cite{Zheng2025}, enhance robustness under high-scattering scenarios but still exhibit domain dependence when confronted with unseen water bodies.

To overcome the limitations of convolution-only models, modern architectures leverage global attention mechanisms. The U-Shape Transformer~\cite{peng2023} captures long-range dependencies for stronger contrast recovery but may introduce oversaturation under rapidly varying illumination. Lightweight transformer variants such as Phaseformer~\cite{khan2025} utilize the Fourier phase component for detail enhancement, while compact ViT models like UIEFormer~\cite{qu2025uieformer} prioritize computational efficiency, albeit with reduced performance on small datasets. Hybrid CNN–Transformer networks~\cite{chen2024hybrid} combine traditional enhancement with contrastive transformer refinement to balance local and global features, and modern re-parameterized U-Net variants~\cite{zhu2025new} replace standard convolutions with efficient fusion blocks to improve speed and stability during enhancement. Together, these advances illustrate the rapid evolution of deep-learning-based underwater image enhancement, while also highlighting the persistent challenges of color fidelity, domain generalization, and radiometric 

Depth maps have emerged as valuable tools for enhancing underwater images by providing crucial 3D information that aids in improving various aspects of image quality \cite{chen2023}. Depth-driven enhancement techniques utilize this depth information to adjust contrast and correct colors. Akkaynak et al.'s SeaThru method \cite{akkaynak2019} significantly enhances clarity by estimating range-dependent attenuation coefficients using range maps generated from external methods like Structure-from-Motion (SfM). The critical flaw here is that errors in SfM or depth estimation directly lead to errors in color and contrast restoration.
Similarly P2CNet~\cite{rao2023} utilizes transmission priors through multi-scale volumetric fusion of texture and color features for probabilistic color compensation. 

Despite the extensive research across traditional, deep learning, and physics-informed models, a gap remains for a robust, unsupervised, and domain-invariant underwater image enhancement architecture capable of maintaining consistent quality across the diverse spectral statistics and turbidity levels found in real-world datasets. Current methods either suffer from poor generalization due to dataset-specific biases or rely on prohibitive paired supervision. To demonstrate the efficacy of our solution, we will rigorously evaluate the proposed DIVER architecture against a comprehensive set of baselines which includes classical, supervised learning methods and unsupervised learning methods.

\section{DIVER}\label{s3}
\noindent

\begin{figure}[h]
     \centering
     \includegraphics[width=0.85\linewidth]{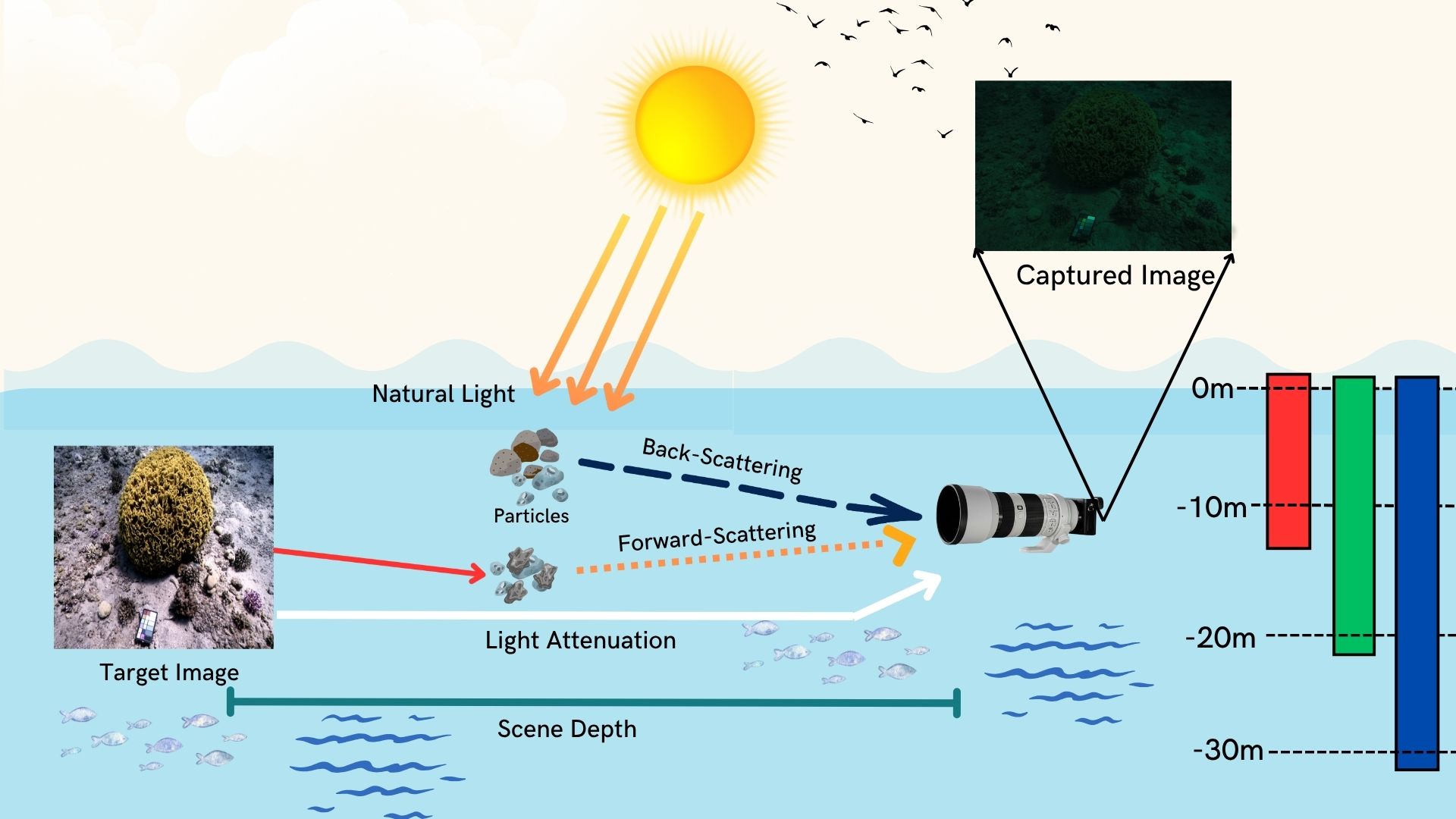}
\caption{Illustration of the underwater image formation process. As natural light penetrates the water column, shorter wavelengths (blue/green) dominate due to wavelength-dependent absorption, while scattering from suspended particles introduces haze and color distortion. The resulting captured image exhibits reduced contrast, spectral imbalance, and optical blur.}
       \label{underwaterimage}
\end{figure}

Underwater imaging is inherently challenging due to the complex interaction between light and the aquatic medium, where different domains exhibit varying water types, lighting conditions, and turbidity levels. As light travels from the surface into deeper waters, it undergoes wavelength-dependent absorption and scattering by suspended particles, leading to severe degradation in color fidelity, contrast, and structural visibility. Figure~\ref{underwaterimage} illustrates this process: sunlight entering the water column is rapidly attenuated, with longer wavelengths such as red vanishing within the first few meters, followed by green and finally blue at greater depths, producing the characteristic blue–green dominance in captured images. Simultaneously, forward scattering caused by particles along the line of sight introduces blur and reduces sharpness, while backscattering redirects ambient light back toward the camera, forming a haze-like veil that washes out scene details. 

As shown in the Figure~\ref{underwaterimage}, the combined effects of light attenuation, depth-dependent spectral loss, and particle-induced scattering transform a clear target scene into a severely degraded captured image, disrupting both global color consistency and local contrast. These distortions—further compounded across domains such as shallow tropical waters, deep oceanic environments, and turbid estuaries, highlighting the need for domain-invariant enhancement methods. In addition, non-uniform illumination from uneven natural or artificial lighting obscures structures in darker regions, and uncertainty in depth estimation complicates attenuation correction. Collectively, these factors lead to luminance loss, reduced sharpness, spectral imbalance, and visibility degradation, making underwater image restoration a fundamentally challenging problem. Next, we mathematically formalize these degradations and present the problem formulation.

\subsection{Problem Formulation}\label{sec:problemformulation}

The multiple intertwined distortions described earlier can be mathematically expressed through the underwater image formation model \cite{akkaynak2019}:
\begin{equation}
U(x) = U_J(x) e^{-\beta z(x)} + B_{\infty} (1 - e^{-\beta z(x)}),
\end{equation}
where \( U(x) \) is the observed image, \( U_J(x) \) is the scene radiance, \( \beta \) is the wavelength-dependent attenuation coefficient, $z(x)$ represents the scene depth, defined as the distance between the imaging sensor (camera) and the object point in the scene, and \( B_{\infty} \) denotes the asymptotic backscatter component from water and particles. The first term models the attenuated direct transmission, while the second term represents the additive backscatter that increases with distance.

The combined effects of wavelength-dependent absorption, particle-induced scattering, and non-uniform illumination cause underwater images to exhibit low contrast, spectral imbalance, structural ambiguity, and significant luminance degradation. These distortions not only affect human perception but also severely hinder downstream computer vision tasks such as object detection, navigation, and mapping. Mathematically, these degradations arise from the underwater image formation model, where the captured image 
$U(x)$ represents a physically degraded version of the true scene radiance $U_
J(x)$ under depth-dependent attenuation and scattering processes. Thus, the enhancement task becomes an inverse problem: estimating 
$U_J(x)$ from its degraded observation $U(x)$ under the constraints imposed by underwater light propagation.

The objective is to reconstruct an image that is both radiometrically consistent and perceptually balanced, while maintaining robust generalization across different water types, lighting conditions, and turbidity levels. In the absence of ground-truth clean images which makes supervised learning impractical, the goal becomes achieving domain-invariant enhancement through unsupervised, physics-guided restoration.
Furthermore, the enhanced output should demonstrate clear improvements on reference and non-reference metrices and exhibit reliable qualitative gains in color fidelity, contrast recovery, and structural clarity under diverse underwater domains, thereby achieving domain-invariant visual enhancement and restoration.

\begin{figure*}[h]
     \centering
     \includegraphics[width =\linewidth, height =0.22  \textheight]{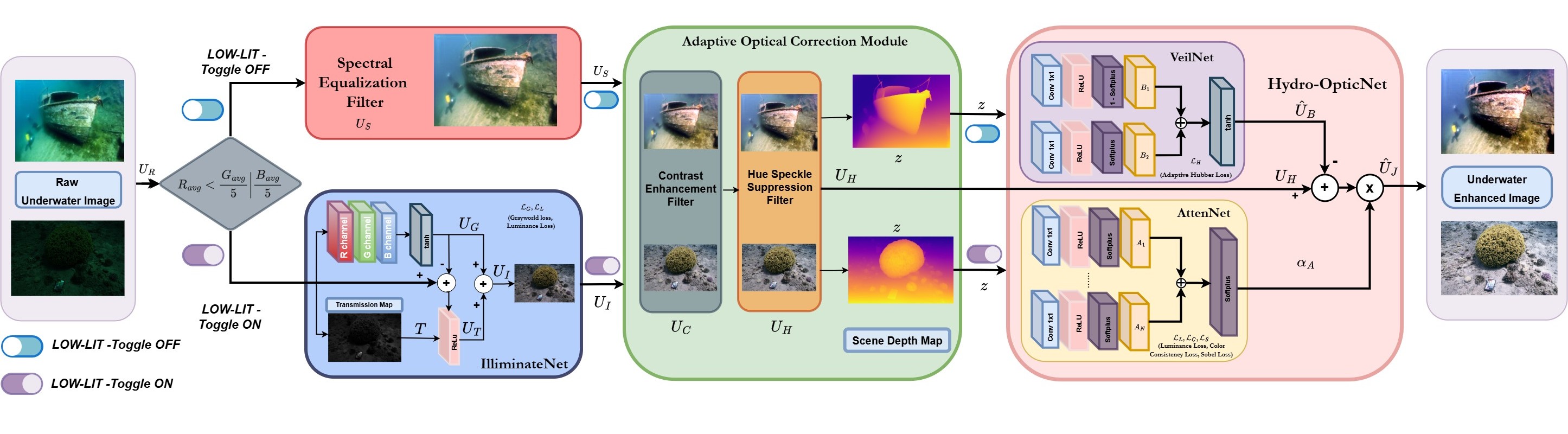}
     \caption{The proposed \textbf{DIVER} architecture for domain-invariant underwater image enhancement. An illumination assessment first determines whether the input image is low-lit. Low-light inputs are routed to the CNN-based IlluminateNet and optimized using the losses $\mathcal{L}_G$ and $\mathcal{L}_L$, while adequately lit images are enhanced through the Spectral Equalization Filter (SEF). The resulting output is further refined by the Adaptive Optical Correction Module (AOCM), which enhances local contrast and suppresses chromatic speckle artifacts. Finally, the Hydro-OpticNet performs physics-guided restoration through VeilNet and AttenNet, which estimate backscatter and wavelength-dependent attenuation, respectively, producing radiometrically consistent results across diverse underwater domains. VeilNet is trained using the adaptive hubber loss $\mathcal{L}_H$, whereas AttenNet employs a composite loss comprising $\mathcal{L}_{L}$, $\mathcal{L}_{C}$, and $\mathcal{L}_{S}$.}
     \label{Joe_arch}
 \end{figure*}

\subsection{DIVER Architecture}\label{sec:diverarchitecture}

The proposed DIVER architecture, an unsupervised and domain-invariant underwater image restoration framework designed to operate reliably across diverse aquatic environments is illustrated in Figure~\ref{Joe_arch}. DIVER unifies empirical enhancement and physics-guided correction within a multi-stage learning framework, enabling robust adaptation across heterogeneous underwater domains—ranging from shallow tropical waters with strong ambient light to deep oceanic scenes and highly turbid estuarine environments.

The process begins with an illumination assessment module, which analyzes the average channel intensities of the input to determine whether it is low-lit or adequately illuminated. This decision routes the image through one of two enhancement paths: Low-lit inputs are processed by IlluminateNet, a CNN-based luminance and color-recovery module. It estimates a depth-guided transmission map and learns global atmospheric light parameters, enabling it to correct illumination falloff, suppress color bias, and produce an achromatic, luminance-balanced output. Losses $\mathcal{L}_G$ and $\mathcal{L}_L$ guide this block to preserve global brightness consistency. For images with sufficient illumination, Spectral Equalization Filter (SEF) performs adaptive spectral normalization by re-balancing the RGB channels to counter wavelength-dependent color shifts. SEF prevents over-correction in well-lit scenes and prepares the image for subsequent contrast enhancement. Both paths converge into the Adaptive Optical Correction Module (AOCM), which consists of two filtering branches: a contrast enhancement filter and a hue-speckle suppression filter. AOCM improves contrast and suppresses chromatic speckles through channel-adaptive filtering. Finally, the \textit{Hydro-OpticNet}, composed of \textit{VeilNet} and \textit{AttenNet}, performs physics-guided estimation of backscatter and wavelength-dependent attenuation using losses $\mathcal{L}_B$, $\mathcal{L}_L$, $\mathcal{L}_C$, and $\mathcal{L}_S$, producing a radiometrically consistent and perceptually balanced enhancement across diverse underwater domains. Each of the blocks in the DIVER architecture is explained in detail in the subsequent sections.


\subsubsection{Illumination Assessment}
As shown in Figure~\ref{Joe_arch}, the enhancement process begins with an illumination assessment, where a simple luminance-based criterion determines whether the input image is low-lit. Specifically, if the average red-channel intensity satisfies  
\[
R_{avg} < \frac{G_{avg}}{5} \;\; \text{or} \;\; R_{avg} < \frac{B_{avg}}{5},
\]  
the image is classified as low-light and routed to \textit{IlluminateNet}, which performs luminance recovery and color rebalance. Otherwise, the image is treated as adequately illuminated and passed directly to the \textit{Spectral Equalization Filter (SEF)} for channel normalization.

\subsubsection{Spectral Equalization Filter (SEF)}

When the illumination assessment determines that the underwater image is well-lit, it is processed by the Spectral Equalization Filter (SEF) for channel neutralization. SEF compensates for wavelength-dependent spectral imbalance, particularly the rapid attenuation of red light. In turbid waters, the green channel often dominates, as seen in OceanDark~\cite{marques2020}.
SEF identifies the dominant channel and uses it as a spectral reference to reinforce weaker ones. Let $\mu^c$ represent the mean intensity for channel $c \in {R,G,B}$ and $C_{\mathrm{sup}} = \max(\mu^R, \mu^G, \mu^B)$ denote the superior channel mean, the relative gain is defined as 
\begin{equation}
g_c = \tfrac{C_{\mathrm{sup}}}{\mu^c + \varepsilon}
\end{equation}
and each channel is normalized using,
\begin{equation}
    U_{S}^c(x) = U^c(x)(g_c)^{\alpha}, 
\end{equation}
where $x$ represents pixel and $\alpha \in (0,1]$ controls the correction strength. The SEF output produces a color-neutral, achromatic image that is subsequently refined in the Adaptive Optical Correction Module (AOCM).

\subsubsection{IlluminateNet}
In underwater environments, longer wavelengths such as red attenuate much faster than green and blue, resulting in significant luminance loss and color imbalance. While the SEF compensates for channel disparities, it becomes less effective when all color channels exhibit low mean intensities, as typically observed in deep waters or poorly lit scenes. To overcome this, IlluminateNet enhances low-light underwater images by restoring intensity and correcting spectral imbalance. In datasets such as SeaThru \cite{akkaynak2019} and FISHTRAC \cite{dawkins2024}, the red channel is particularly diminished relative to green and blue, making IlluminateNet crucial for recovering perceptual brightness and natural color fidelity. To mitigate this, IlluminateNet takes the raw underwater image \({U_R}\) as input, producing a luminance-corrected output $U_I$. The schematic of the IlluminateNet is shown in Figure \ref{Joe_arch}. IlluminateNet integrates a simplified image formation model, 
\begin{equation}
    U_R = U_I T+ U_G(1-T) \label{RS}
\end{equation}
where, $T$ represents transmission map and $U_G$ is the global atmospheric light. Eq. (\ref{RS}) is used to estimate scene illumination $U_I$ as:
\begin{equation}
U_I = U_G + U_T
\label{enhace}
\end{equation}
where, 
  \begin{equation}
 U_T=\frac{(U_R-U_G)}{T}
 \end{equation}
The transmission map $T$ is calculated based on the difference between the underwater image and the corresponding atmospheric light. It is computed as:
\begin{equation}
T(x) = \max_{x \in \psi(y)} \left( \frac{| U_R(x) - A |}{\max(A, 1 - A)} \right) \label{trans}
\end{equation}
where \( \psi(y) \) denotes a small local patch and $A$ is the ambient light. $A$ is computed as the mean RGB intensity of the top 0.1\% of pixels corresponding to the farthest regions in the estimated depth \cite{makam2025}. The global light component $U_G$ is the learnable parameter of IlluminateNet.It is estimated by independently processing the R, G, and B channels through separate convolutional networks with ReLu $\mathcal{R}$, followed by concatenation and a tanh activation function $\mathcal{T}$. 
\begin{equation}
    U_G = \mathcal{T}(\mathcal{R}(R) \odot \mathcal{R}(G) \odot \mathcal{R}(B)),
\end{equation}
where, $\odot$ represents concatenation.  This separation allows channel-wise learning of wavelength-dependent attenuation effects, ensuring accurate compensation for spectral variations \cite{makam2025}. The tanh activation $\mathcal{T}$ function is used to bound the channels to $(-1,1)$, enabling both positive and negative gray-level adjustments. Negative gray levels are interpreted as light intensifiers, enabling the dynamic range of a low-light image to be extended across the full grayscale spectrum \cite{jourlin2024}. This bounded activation prevents excessive radiance amplification or instability caused by steep gradients, and preventing numerical overflow.

Further, the transmission map $T$ calculated using \eqref{trans} is passed through a convolution layer with ReLu activation. The resulting feature is combined with $(U_R - U_G)$ to form $U_T$. It is then added to  $U_G$  to generate the final luminance-corrected output $U_I$. 
\begin{equation}
    U_I = U_G + \mathcal{R}\left(\frac{U_R - U_G}{T}\right)
\end{equation}
The IlluminateNet output $U_I$ is an achromatic and intensity-balanced image, providing luminance regardless of whether the scene originates from shallow or deep waters. In this block, the only learnable parameter is the global atmospheric light, which is optimized using the loss functions described in Section~\ref{lossIN}.

While IlluminateNet for low-lit images or SEF for adequately lit images effectively corrects non-uniform illumination and rebalances the RGB channels, it does not explicitly enhance hue, saturation or local contrast. This motivates the introduction of the AOCM block, which further balances colorfulness, hue and contrast to produce a perceptually enriched intermediate representation before physics-guided restoration.

\subsubsection{Adaptive Optical Correction Module (AOCM)}

After color rebalancing in either \textit{IlluminateNet} or the \textit{SEF} branch of the DIVER architecture, the resulting achromatic image is forwarded to the Adaptive Optical Correction Module (AOCM). This module applies a sequence of corrections to enhance contrast, suppress hue-related speckling, and recover structural clarity under diverse underwater conditions. AOCM operates through two key components: the \textit{Contrast Enhancement Filter (CEF)} and the \textit{Hue Speckle Suppression Filter (HSSF)}, which together refine global appearance and stabilize chromatic distortions.

\textbf{Contrast Enhancement Filter (CEF):}
Following spectral equalization, CEF addresses contrast improvement by redistributing pixel intensities adaptively across bright and dark regions \cite{azmi2019}. CEF restores visibility by independently stretching low- and high-intensity regions. This effectively recover lost dynamic range and reduce the hazy appearance of underwater images. For each channel, the mean ($\mu^c$) and median ($\tilde{m}^c$) intensity are combined to define a pivot point ($t^c$) that divides the luminance range:
\begin{align}
t^c &= \tfrac{1}{2}\big(\mu^c+\tilde m^c\big),
\end{align}
Pixels below and above this pivot are independently stretched toward the lower and upper bounds using piecewise mapping - dual stretching:
\begin{eqnarray}
    U^c_{\mathrm{LS}}(x)\!\!&=&\!\!
\begin{cases}
(x-m^c)\,\dfrac{255-m^c}{t^c-m^c}+m^c, \!\!\!\!\!\! &x<t^c,\\[4pt]
255, & \!\!\! x\ge t^c,
\end{cases}\quad \\
U^c_{\mathrm{US}}(x) \!\!& = &\!\!
\begin{cases}
0,&x<t^c,\\[4pt]
(x-t^c)\,\dfrac{255}{M^c-t^c},&x\ge t^c.
\end{cases}
\end{eqnarray}
where,
\begin{eqnarray}
m^c = \min(U^c_{S}), & \quad M^c = \max(U^c_{S}), \nonumber \\
\mu^c = \mathrm{mean}(U^c_{S}), & \quad \tilde{m}^c = \mathrm{median}(U^c_{S})
\end{eqnarray}
and $U^c_{S} \in [0,255]^{H\times W}$. The two luminance‐stretched images obtained from CEF corresponding to low and high intensity enhancements are integrated to form a balanced exposure image:
\begin{eqnarray}
U_{C}^c=\tfrac{1}{2}\big(U^c_{\mathrm{LS}}(x)+U^c_{\mathrm{US}}(x)\big), 
\end{eqnarray}
The fused output ensures illumination-invariant brightness, ensuring contrast enhancement addressing shadowed regions while preserving fine details. 

\textbf{Hue Speckle Suppression Filter (HSSF)}:
Suspended particles and floating matter often create high‐saturation color speckles, predominantly within the cyan–aqua hue range. HSSF mitigates these chromatic artifacts by operating in HSV and CIELAB space.
Define a hue-selective binary mask:
\begin{equation}
M(x) \!\!=\!\!
\begin{cases}
1, \!\!\!\!\!& h_l \le H(x) \le h_u, S(x)\!\ge\!S_{\min},V(x)\!\ge\!V_{\min},\\[3pt]
0, \!\!\!\!\!& \text{otherwise}.
\end{cases}
\label{eq:mask}
\end{equation}
where $H(x), S(x), V(x)$ are the hue, saturation, and value of $x$ in HSV space. 
Let $U_{CEF}$ be converted from RGB to CIELAB color space with three channels: \(L\) (lightness), \(a\) (green–red), and \(b\) (blue–yellow). Now, for $x_{HSV}$ satisfying the binary mask $M(x)=1$, hue suppression is applied by attenuating the $\{a,b\}$ in CIELab space components toward the neutral point $128$. The new values are calculated as: 
\begin{equation}
a'(x_{Lab}) = 128 + (a(x_{Lab})-128)(1-\lambda), 
\end{equation}
\begin{equation}
b'(x_{Lab}) = 128 + (b(x_{Lab})-128)(1-\lambda), \label{eq:chroma}
\end{equation}
where $\lambda$ is the chroma suppression factor.
The hue-corrected image, $U_{HS}(x) \in (L(x),\, a'(x),\, b'(x))$ is converted back to the RGB domain, thus reducing chromatic speckles without affecting true object color, resulting in a visually cleaner image $U_{H}$.
The AOCM ensures spectral and contrast normalization across domains with varying color biases, enabling consistent color reproduction in both blue-dominant and green-tinted waters. The final correction based on physics of underwater image formation is carried out at in the Hydro-OpticNet.

\subsubsection{Hydro-OpticNet}
 Hydro-OpticNet is designed to model and correct the physics degradations that persist after illumination and optical correction. While IlluminateNet which relies on physics to effectively restore luminance, it does not fully capture the nonlinear light decay occurring at greater depths or in highly scattering waters. To address this, Hydro-OpticNet employs the revised underwater image formation model~\cite{akkaynak2019}:
\begin{equation}
    U = U_J U_A + U_{B} \label{eimgfor}
\end{equation}
where \( U \) is the observed degraded image (here $U$ is the output of the AOCM, $U_{H}$), \( U_J \) represents true scene radiance, \( U_A \) models wavelength-dependent attenuation, and \( U_{B} \) denotes the backscatter component arising from reflections by suspended particles that form a light veil in the captured image. The input to Hydro-OpticNet is the output image obtained from the AOCM block $U_{H}$. Hydro-OpticNet comprises two submodules: VeilNet and AttenNet. VeilNet estimates and removes the additive backscatter component caused by suspended particles and illumination haze. AttenNet compensates for wavelength-dependent attenuation to restore true radiance and color fidelity, maintaining stability under uncertain or noisy depth estimates.

\textbf{VeilNet:} The backscattering caused by suspended particles in water increases nonlinearly with scene depth \(z\), eventually saturating into a bluish-teal veil that severely reduces visibility. VeilNet is designed to explicitly estimate and remove this additive backscatter component by integrating a physics-guided representation of the underwater image formation process. The backscatter contribution \(U_B\) is modeled as:
\begin{equation}
    U_B = f_B(z), 
\end{equation}
where
\begin{equation}
    f_B(z) = B_1(1 - e^{-b_1 z}) + B_2 e^{-b_2 z}, \label{eq:veil_physics}
\end{equation}
with \(B_1, B_2\) and \(b_1, b_2\) being learnable parameters that respectively define the asymptotic and exponential decay terms controlling the  backscatter intensity.

Scene depth \(z\) is estimated using the transformer-based monocular model DepthAnythingV2~\cite{yang2024v2}, which demonstrates better cross-domain generalization under underwater color distortions compared to MiDaS~\cite{Ranftl2020} and MonoDepth2~\cite{Godard2019}. Its large-scale pretraining on diverse visual domains enables robust depth inference even under scattering and turbidity variations.

Within Hydro-OpticNet, VeilNet processes the estimated scene depth of the AOCM output \(U_{HASSF}\) through a sequence of convolutional layers with ReLU and Softplus activations to derive the backscatter coefficients \(\hat{b}_1\) and \(\hat{b}_2\). Each exponential component in (\ref{eq:veil_physics}) is implemented as:
\begin{equation}
    \hat{U}_B = \mathcal{T}\left(\hat{B}_1 \, \mathcal{S_C^+}(-\hat{b}_1 z) + \hat{B}_2 \, \mathcal{S^+}(-\hat{b}_2 z)\right),
    \label{eq:veil_nn}
\end{equation}
where \(\mathcal{S^+}(\cdot)\) and \(\mathcal{S_C^+}(\cdot)\) denote the softplus activation and its complementary form, respectively:
\begin{equation}
  \mathcal{S^+}(x) =
  \begin{cases}
    1, & x \leq 0,\\
    \log(1 + e^{-x}), & x > 0,
  \end{cases}
\end{equation}
\begin{equation}
    \mathcal{S_C^+}(x) = 1 - \mathcal{S^+}(x).
\end{equation}

The softplus activation is used instead of the exponential function to maintain numerical stability and prevent gradient explosion during training~\cite{dubey2022}. Its smooth differentiability allows for learning under varying backscatter magnitudes and lighting conditions, while retaining the physical interpretability of exponential attenuation. The outputs from both branches in (\ref{eq:veil_nn}) are summed and passed through a tanh activation function $\mathcal{T}$ to produce the estimated backscatter map \(\hat{U}_B\). The direct signal is then recovered as:
\begin{equation}
    \hat{U}_D = U_{H} - \hat{U}_B.
\end{equation}
This residual learning process allows VeilNet to isolate the direct radiance component effectively, even under severe turbidity, thereby restoring image visibility and color contrast.

\textbf{AttenNet:} The AttenNet block models the wavelength-dependent attenuation with scene depth \(z\), which causes progressive loss of contrast and spectral fidelity in underwater images. AttenNet learns the inverse attenuation function to recover the true radiance component from the direct signal. The attenuation model is expressed as:
\begin{equation}
U_A(z) = f_A(z),
\label{eatt}
\end{equation}
where
\begin{equation}
f_A(z) = \sum_{p = 1}^{P} A_p \exp(-a_p z),
\end{equation}
with \(P\) denoting the number of exponential terms approximating wavelength-dependent decay.

Rather than estimating \(f_A(z)\) directly, AttenNet learns its inverse \((U_A)^{-1} = \alpha_A(z)\) to perform deattenuation. The inverse attenuation mapping is expressed as:
\begin{equation}
\alpha_A(z) = \mathcal{S^+}\left(\sum_{p=1}^{P} \alpha'_p \, \mathcal{S^+}(\alpha_p z)\right),
\end{equation}
where \(\mathcal{S^+}\) again denotes the softplus activation, ensuring smooth estimation of attenuation coefficients. This is illustrated in AttenNet block in Figure~\ref{Joe_arch}. The final restored radiance is obtained as:
\begin{equation}
\hat{U}_{J} = (U_{H} - \hat{U}_B) \, \alpha_A(z).
\end{equation}
By combining the outputs of VeilNet and AttenNet, Hydro-OpticNet produces a radiometrically consistent and perceptually balanced estimate of the scene radiance \(\hat{U}_J\). The module achieves domain invariance by explicitly modeling both the additive backscatter component and the multiplicative wavelength-dependent attenuation within a unified physics-constrained learning, and enabling restoration across diverse water types, depths, and scattering conditions. In the following section, we detail the loss functions that guide the learnable parameters of Hydro-OpticNet for estimation of backscatter and attenuation in Section~\ref{lfvn} and \ref{lfan}.

\subsection{Learning Algorithm} \label{sec:diverlearning}
The DIVER architecture employs dedicated loss functions across its learning-based blocks (IlluminateNet, VeilNet, and AttenNet) to ensure robust illumination recovery, accurate backscatter removal, and faithful radiance restoration. Each loss component is specifically  addresses distinct underwater degradations such as non-uniform illumination, wavelength-dependent attenuation, and structural blurring.

\subsubsection{IlluminateNet Loss Functions} \label{lossIN}
IlluminateNet leverages grayworld and luminous losses to correct illumination and enforce chromatic consistency. 
\begin{itemize}
\item \textbf{Grayworld Loss} \((\mathcal{L}_{G}\)):
Based on the gray-world assumption, this loss enforces chromatic neutrality by constraining the mean of each RGB channel to converge toward a global gray reference. It reduces the dominance of blue or green hues prevalent in underwater scenes and stabilizes the overall white balance. It computes the mean intensity of each RGB channel in the enhanced image and derives a gray reference \(\mu^{g}\)  as the average of these means. The loss is defined as the mean absolute deviation of each channel's mean from \(\mu^{g}\):
\begin{equation}
\mathcal{L}_{G} = \frac{1}{3} \sum_{c \in \{R, G, B\}} \left| \mu^c(U_I) - \mu^{g}(U_I) \right|,
\end{equation}
where \(\mu^c\) is the mean of channel \(c\), and \(\mu^{g} = \frac{1}{3}(\mu^R + \mu^G + \mu^B)\) of IlluminateNet ouput $U_I$.

\item \textbf{Luminous Loss} (\(\mathcal{L}_{L}\)):  
This term preserves global luminance levels to prevent under- or over-exposure during intensity correction. It ensures perceptual brightness uniformity even in low-illumination regions:
\begin{equation}
 \mathcal{L}_{L} = \frac{1}{3} \sum_{c=1}^{3} \left( \frac{1}{N} \sum_{i=1}^{N} ({U_I}_i^c - {U}_{t}^c)^2 \right),
\end{equation}
here \( {U_I}_i^c \) represents the intensity of the \( i \)-th pixel in channel \( c \), \( {U}_{t}^c \) is the target intensity for channel \( c \), which can be a predefined reference value (e.g., 128 (or 0.5) for mid-gray or 255 (or 1.0) for white) and \( N \) is the total number of pixels in the image. 
\item The combined loss stabilizes illumination recovery and enforces chromatic uniformity:
\begin{equation}
\mathcal{L}_{T} = \lambda_1 \mathcal{L}_{G} + \lambda_2 \mathcal{L}_{L},
\end{equation}
where \(\lambda_1 = 0.25\) and \(\lambda_2 = 1\). Thus minimizing this loss the network to produce images with chromatically consistent output by reducing color biases across channels.
\end{itemize}

\subsubsection{VeilNet Loss Function} \label{lfvn}
VeilNet applies the adaptive huber loss to robustly estimate and remove the effects of backscatter. The Adaptive Huber Loss (\(\mathcal{L}_{H}\))~\cite{sun2020} is used to train VeilNet, formulated as:
\begin{equation}
  \mathcal{L}_{H} =
  \begin{cases}
      (\hat{U}_B)^2, & \text{if } |\hat{U}_B| \leq \delta, \\
      \eta \delta \, (|\hat{U}_B| - \tfrac{\delta}{2}), & \text{otherwise},
  \end{cases}
\end{equation}
where \(\delta\) defines the transition threshold between quadratic and linear regimes, and \(\eta\) balances their relative contributions. 
By adjusting its penalization regime, the Adaptive Huber Loss stabilizes training under varying turbidity and depth conditions. It effectively suppresses the influence of outlier pixels caused by strong scattering, while still maintaining fine-grained sensitivity to subtle radiance variations. This makes it well-suited for estimating the backscatter veil, which exhibits nonlinear growth with scene depth.

\subsubsection{AttenNet Loss Functions}\label{lfan}
To guide the learning of attenuation correction, a composite loss function is employed that balances luminance fidelity, chromatic consistency, and structural detail preservation. AttenNet refines the image by integrating luminance, color consistency, and sobel edge losses.

\begin{itemize}

\item \textbf{Luminance Loss} (\(\mathcal{L}_{L}\)):  
Although used earlier in IlluminateNet, its inclusion here serves a different purpose. In IlluminateNet, it restores global exposure under uneven lighting, whereas in AttenNet, it stabilizes radiance recovery across depths, compensating for exponential light decay (\(U \propto e^{-\beta z}\)). This prevents overcompensation during inverse attenuation learning and ensures that brightness remains physically consistent with estimated depth.

\item \textbf{Color Consistency Loss} (\(\mathcal{L}_{C}\)):  
While conceptually related to Grayworld, this term acts locally rather than globally, enforcing inter-channel balance within spatial neighborhoods rather than over the entire image:
\begin{equation}
\mathcal{L}_{C} = \sum_{(p,q) \in c'} \left( \hat{U}_J^p - \hat{U}_J^q \right)^2,
\end{equation}
where \( c' = \{(R, G), (R, B), (G, B)\} \), and \( \hat{U}_J^p \) denotes the mean intensity of channel \( p \). 
This refinement is critical for attenuation correction, as wavelength-dependent decay varies spatially with scene depth. Thus, \(\mathcal{L}_{CC}\) ensures that local chromatic distortions (e.g., red loss in deeper pixels) are harmonized, yielding spectrally balanced radiance.

\item \textbf{Sobel Edge Loss} ($\mathcal{L}_{S}$) : To counteract optical blur induced by multiple scattering, a gradient-based loss maintains edge and structural integrity. Sobel-based gradient consistency loss compares edges between the reconstructed image \( \hat{U}_J^c \) and the direct signal \( \hat{U}_D^c \):
      \begin{align}
     \mathcal{L}_{S} & = \frac{1}{3} \sum_{c} \left( \left| \mathbf{S_x} * \hat{U}^c_{J} - \mathbf{S_x} * \hat{U}^c_{D} \right| \right. \nonumber \\
     & \quad + \left. \left| \mathbf{S_y} * \hat{U}^c_{J} - \mathbf{S_y} * \hat{U}^c_{D} \right| \right),
\end{align}
where \( * \) denotes convolution, and \( \mathbf{S_x} \) and \( \mathbf{S_y} \) are the standard Sobel filters:
\begin{equation}
\mathbf{S_x} =
\begin{bmatrix}
1 & 0 & -1 \\
2 & 0 & -2 \\
1 & 0 & -1
\end{bmatrix}, \quad
\mathbf{S_y} =
\begin{bmatrix}
1 & 2 & 1 \\
0 & 0 & 0 \\
-1 & -2 & -1
\end{bmatrix}.
\end{equation}
This ensures that fine textural features and object boundaries degraded by scattering are preserved in the restored output.
\item The total loss is given by:
\begin{equation}
\mathcal{L}_{A} = \mathcal{L}_{L} + \mathcal{L}_{C} + \mathcal{L}_{S}.
\end{equation}
\end{itemize}
The composite loss design promotes domain invariance by jointly optimizing luminance fidelity, chromatic balance, and structural preservation across varied underwater domains, ensuring that the learned representation generalizes beyond dataset-specific lighting or spectral biases.

In summary, the DIVER architecture addressed the challenges through its cascaded modules. \textit{IlluminateNet} corrects low-light conditions and normalizes luminance to counter depth-dependent radiance loss. For adequately lit scenes, the \textit{SEF} equalizes dominant spectral channels, mitigating wavelength-selective attenuation. The \textit{AOCM} further enhances global contrast and suppresses hue speckles caused by scattering and uneven local illumination. Finally, \textit{Hydro-OpticNet} (VeilNet and AttenNet) performs physics-guided correction by removing backscatter haze and compensating for wavelength-dependent attenuation governed by scene depth. Together, these components jointly resolve luminance decay, spectral imbalance, optical blur, and scattering-induced haze, enabling a domain-invariant restoration of underwater images across varied and challenging environments.

\section{Performance evaluation of DIVER }\label{s4}
In this section, we present a comprehensive evaluation of the proposed DIVER architecture. We first describe the datasets used for the evaluation, the evaluation metrics, and implementation details used throughout our study. We then conduct both quantitative and qualitative comparisons against ten state-of-the-art underwater image enhancement methods across eight publicly available datasets, covering paired and unpaired testing scenarios, ablation study on the performance of individual blocks of DIVER . Finally, we assess the impact of DIVER on downstream robotic perception tasks using SIFT feature analysis, demonstrating its effectiveness under challenging real-world underwater conditions.

\subsection{Datasets Used in this Study}

The datasets used in this study span a broad spectrum of underwater imaging conditions, ensuring that DIVER is evaluated across diverse spectral, illumination, and visibility regimes. SeaThru~\cite{akkaynak2019} offers paired data with range information, capturing natural underwater color casts and illumination drop-off with depth. OceanDark~\cite{marques2020} represents extremely low-light environments with strong shadows and non-uniform illumination due to artificial light during deep-sea operations. The USOD10K dataset~\cite{Hong2025} consists of images exhibiting severe color distortion, suspended particle scattering, and spatially varying visibility conditions common in turbid and coastal waters. While U45~\cite{drews2013} focuses on low-contrast, haze-dominated scenes, FISHTRAC~\cite{dawkins2024} includes challenging scenes where the red channel is heavily attenuated relative to green and blue, producing a characteristic cyan-green cast. 

The UIEB dataset~\cite{li2019} contains real-world underwater scenes paired with high-quality reference images, covering bluish and greenish color dominance, varying illumination, and heterogeneous water types. The UFO-120 dataset~\cite{wang2019deep} includes images acquired across multiple locations with different turbidity levels and lighting conditions, enabling evaluation under spatially varying backscatter and attenuation. EUVP~\cite{islam2020} provides both paired and unpaired subsets, featuring a mixture of low-light, well-lit, hazy, and clear-water scenarios suitable for learning illumination- and haze-invariant features. LSUI~\cite{peng2023} further expands this diversity with scenes captured in coastal, deep-sea, and cave environments, introducing strong color casts and complex lighting variations. 
Collectively, these datasets encompass diverse water types, depth ranges, turbidity levels, illumination conditions, and spectral distortions. This diversity allows a comprehensive and domain-general evaluation of the DIVER architecture across both controlled and real-world underwater environments.

\subsection{Performance Evaluation Metrics}

To quantitatively evaluate the performance of the proposed DIVER architecture, both reference-based and non-reference based metrics are employed. The reference-based metrics assess similarity with respect to ground-truth images, whereas non-reference metrics evaluate perceptual quality when ground truth is unavailable an essential consideration in underwater imaging.

\subsubsection{Reference-Based Metrics}
Reference-based metrics evaluate restoration accuracy by comparing the enhanced output against ground-truth clean images. These metrics provide objective measures of fidelity, structural preservation, and perceptual similarity, enabling fair comparison across enhancement methods.

\begin{itemize}
    \item Peak Signal-to-Noise Ratio (PSNR)  \cite{panetta2015,li2019}:
PSNR quantifies image fidelity by comparing the reconstructed image with its ground truth based on pixel-wise intensity differences. A higher PSNR value indicates better signal preservation and lower reconstruction error. 
\item Structural Similarity Index (SSIM): SSIM, on the other hand, measures perceptual similarity by evaluating luminance, contrast, and structural correspondence between the two images. Higher SSIM values signify improved retention of textural and structural details, offering a perceptually meaningful measure of restoration quality. 
\end{itemize}

\subsubsection{Non-Reference Metrics}
Non-reference metrics assess the quality of underwater images without requiring ground-truth references, making them essential for evaluating real-world scenarios where clean images are unavailable. These metrics quantify perceptual attributes such as color balance, contrast, sharpness, and naturalness factors that are strongly affected by underwater optical degradation.

\begin{itemize}
    \item Gray Patch Mean Angular Error (GPMAE) \cite{berman2017}: 
        GPMAE measures the deviation of pixel chromaticity from an ideal gray reference based on the assumption that well-balanced underwater images contain neutral patches with equal RGB contributions. It is computed as:
         \begin{equation}
            \bar{\psi} = \frac{1}{6} \sum_{x_i} \cos^{-1} \left( \frac{\mathbf{U}(x_i) \cdot (1, 1, 1)}{\|\mathbf{U}(x_i)\| \cdot \sqrt{3}} \right),
        \end{equation}
        where \( x_i \) are the coordinates of gray patches. Lower angular errors indicate better color correction and reduced chromatic bias.

    \item Underwater Image Quality Measure (UIQM) \cite{panetta2015}: 
        UIQM is a widely used no-reference metric combining colorfulness, sharpness, and contrast to estimate perceptual quality in underwater imagery. It is defined as:
        \begin{equation}
        \text{UIQM} = c_1 \times \text{UICM} + c_2 \times \text{UISM} + c_3 \times \text{UIConM},
        \end{equation}
         where \( \text{ UICM (Underwater Image Colorfulness Measure)} \), \( \text{UISM (Underwater Image Sharpness Measure)} \), and \( \text{UIConM (Underwater Image Contrast Measure)} \) denote the colorfulness, sharpness, and contrast components, respectively, and \( c_1 = 0.0282 \), \( c_2 = 0.2953 \), \( c_3 = 3.5753 \) are empirically determined coefficients. Higher UIQM values correspond to images with better perceptual quality, improved contrast, and enhanced color fidelity.

\item Underwater Color Image Quality Evaluation (UCIQE) \cite{Yang2015}:
UCIQE is a non-reference metric designed to assess underwater image quality by combining chroma, saturation, and contrast components:
\begin{equation}
\text{UCIQE} = c_1 \times \text{Chroma} + c_2 \times \text{Saturation} + c_3 \times \text{Contrast},
\end{equation}
where \( \text{Chroma} \), \( \text{Saturation} \), and \( \text{Contrast} \) represent image attributes measured in the CIELab color space, and \( c_1 \), \( c_2 \), \( c_3 \) are empirically determined coefficients. Higher UCIQE values indicate better underwater image quality.

\item Blind/Referenceless Image Spatial QUality Evaluator (BRISQUE)  \cite{mittal2011}: BRISQUE is a no-reference metric for evaluating Natural Scene Statistics. It analyzes locally normalized pixel intensities and their spatial relationships, models their statistical behavior, and uses these features to estimate how much naturalness has been lost due to artifacts. A lower BRISQUE score indicates a more natural-looking and higher-quality image.

\end{itemize}

\begin{figure*}[htbp]
    \centering

    \setlength{\tabcolsep}{0pt}
    \renewcommand{\arraystretch}{0}

\begin{tabular}{@{}
    >{\centering\arraybackslash}m{0.9cm}
    @{\hspace{0.3em}}*{3}{@{}c@{}} 
    @{\hspace{1em}}               
    *{3}{@{}c@{}}                 
    @{\hspace{1em}}               
    *{3}{@{}c@{}}@{}}
        
        & \multicolumn{3}{c}{\textbf{(a)~\textit{SeaThru}}} &
          \multicolumn{3}{c}{\textbf{(b)~\textit{OceanDark}}} &
          \multicolumn{3}{c}{\textbf{(c)~\textit{USOD10K}}} \\[0.3em]  
        \textbf{\scriptsize\textit{Raw}} &
        \includegraphics[width=0.1\textwidth,height=0.1\textwidth]{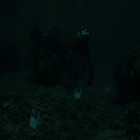} &
        \includegraphics[width=0.1\textwidth,height=0.1\textwidth]{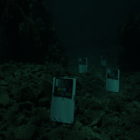} &
        \includegraphics[width=0.1\textwidth, height=0.1\textwidth]{JournalFigures/Comparisions/ST/Raw/LFT_3383.png} &
        \includegraphics[width=0.1\textwidth, height=0.1\textwidth]{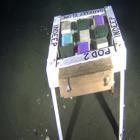} &
        \includegraphics[width=0.1\textwidth, height=0.1\textwidth]{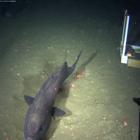} &
        \includegraphics[width=0.1\textwidth, height=0.1\textwidth]{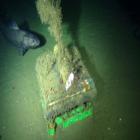} &
        \includegraphics[width=0.1\textwidth, height=0.1\textwidth]{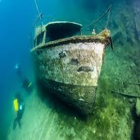} &
        \includegraphics[width=0.1\textwidth, height=0.1\textwidth]{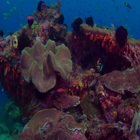} &
        \includegraphics[width=0.1\textwidth, height=0.1\textwidth]{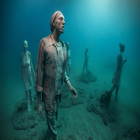} \\[-0.1em]
        
        \textbf{\scriptsize\textit{IBLA}} &
        \includegraphics[width=0.1\textwidth,height=0.1\textwidth]{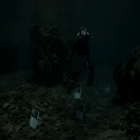} &
        \includegraphics[width=0.1\textwidth,height=0.1\textwidth]{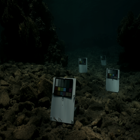} &
        \includegraphics[width=0.1\textwidth, height=0.1\textwidth]{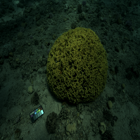} &
        \includegraphics[width=0.1\textwidth, height=0.1\textwidth]{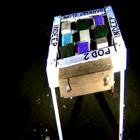} &
        \includegraphics[width=0.1\textwidth, height=0.1\textwidth]{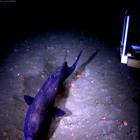} &
        \includegraphics[width=0.1\textwidth, height=0.1\textwidth]{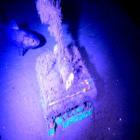} &
        \includegraphics[width=0.1\textwidth, height=0.1\textwidth]{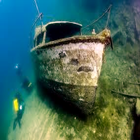} &
        \includegraphics[width=0.1\textwidth, height=0.1\textwidth]{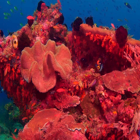} &
        \includegraphics[width=0.1\textwidth, height=0.1\textwidth]{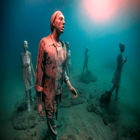} \\[-0.1em]

        \textbf{\scriptsize\textit{DCP}} &
        \includegraphics[width=0.1\textwidth, height=0.1\textwidth]{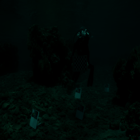} &
        \includegraphics[width=0.1\textwidth, height=0.1\textwidth]{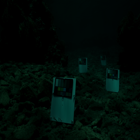} &
        \includegraphics[width=0.1\textwidth, height=0.1\textwidth]{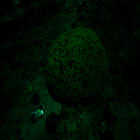} &
        \includegraphics[width=0.1\textwidth, height=0.1\textwidth]{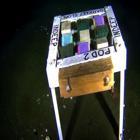} &
        \includegraphics[width=0.1\textwidth, height=0.1\textwidth]{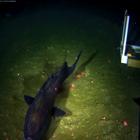} &
        \includegraphics[width=0.1\textwidth, height=0.1\textwidth]{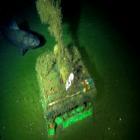} &
        \includegraphics[width=0.1\textwidth, height=0.1\textwidth]{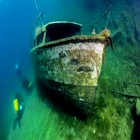} &
        \includegraphics[width=0.1\textwidth, height=0.1\textwidth]{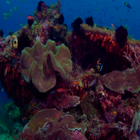} &
        \includegraphics[width=0.1\textwidth, height=0.1\textwidth]{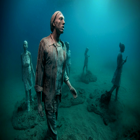} \\[-0.1em]

        \textbf{\scriptsize\textit{UDCP}} &
        \includegraphics[width=0.1\textwidth, height=0.1\textwidth]{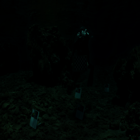} &
        \includegraphics[width=0.1\textwidth, height=0.1\textwidth]{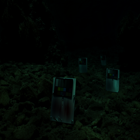} &
        \includegraphics[width=0.1\textwidth, height=0.1\textwidth]{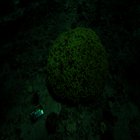} &
        \includegraphics[width=0.1\textwidth, height=0.1\textwidth]{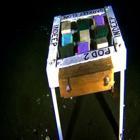} &
        \includegraphics[width=0.1\textwidth, height=0.1\textwidth]{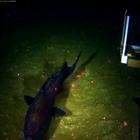} &
        \includegraphics[width=0.1\textwidth, height=0.1\textwidth]{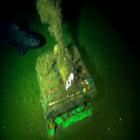} &
        \includegraphics[width=0.1\textwidth, height=0.1\textwidth]{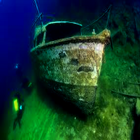} &
        \includegraphics[width=0.1\textwidth, height=0.1\textwidth]{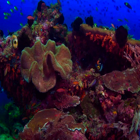} &
        \includegraphics[width=0.1\textwidth, height=0.1\textwidth]{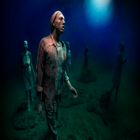} \\[-0.1em]

        \textbf{\scriptsize\textit{ULAP}} &
        \includegraphics[width=0.1\textwidth, height=0.1\textwidth]{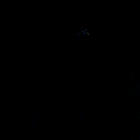} &
        \includegraphics[width=0.1\textwidth, height=0.1\textwidth]{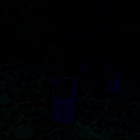} &
        \includegraphics[width=0.1\textwidth, height=0.1\textwidth]{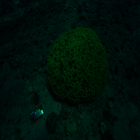} &
        \includegraphics[width=0.1\textwidth, height=0.1\textwidth]{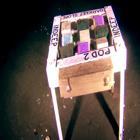} &
        \includegraphics[width=0.1\textwidth, height=0.1\textwidth]{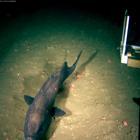} &
        \includegraphics[width=0.1\textwidth, height=0.1\textwidth]{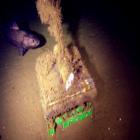} &
        \includegraphics[width=0.1\textwidth, height=0.1\textwidth]{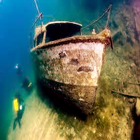} &
        \includegraphics[width=0.1\textwidth, height=0.1\textwidth]{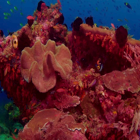} &
        \includegraphics[width=0.1\textwidth, height=0.1\textwidth]{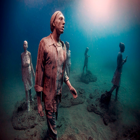} \\[-0.1em]

        \textbf{\scriptsize\textit{P2CNet}} &
        \includegraphics[width=0.1\textwidth, height=0.1\textwidth]{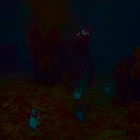} &
        \includegraphics[width=0.1\textwidth, height=0.1\textwidth]{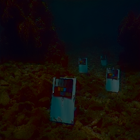} &
        \includegraphics[width=0.1\textwidth, height=0.1\textwidth]{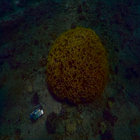} &
        \includegraphics[width=0.1\textwidth, height=0.1\textwidth]{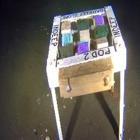} &
        \includegraphics[width=0.1\textwidth, height=0.1\textwidth]{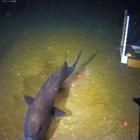} &
        \includegraphics[width=0.1\textwidth, height=0.1\textwidth]{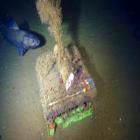} &
        \includegraphics[width=0.1\textwidth, height=0.1\textwidth]{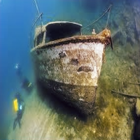} &
        \includegraphics[width=0.1\textwidth, height=0.1\textwidth]{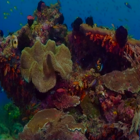} &
        \includegraphics[width=0.1\textwidth, height=0.1\textwidth]{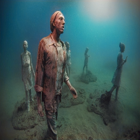} \\[-0.1em]

        \textbf{\scriptsize\textit{PF}} &
        \includegraphics[width=0.1\textwidth, height=0.1\textwidth]{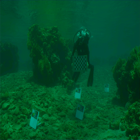} &
        \includegraphics[width=0.1\textwidth, height=0.1\textwidth]{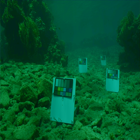} &
        \includegraphics[width=0.1\textwidth, height=0.1\textwidth]{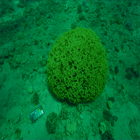} &
        \includegraphics[width=0.1\textwidth, height=0.1\textwidth]{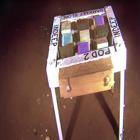} &
        \includegraphics[width=0.1\textwidth, height=0.1\textwidth]{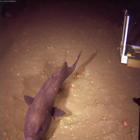} &
        \includegraphics[width=0.1\textwidth, height=0.1\textwidth]{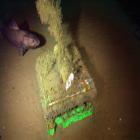} &
        \includegraphics[width=0.1\textwidth, height=0.1\textwidth]{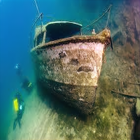} &
        \includegraphics[width=0.1\textwidth, height=0.1\textwidth]{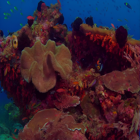} &
        \includegraphics[width=0.1\textwidth, height=0.1\textwidth]{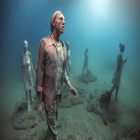} \\[-0.1em]

        \textbf{\scriptsize\textit{WaterNet}} &
        \includegraphics[width=0.1\textwidth, height=0.1\textwidth]{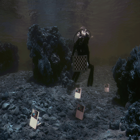} &
        \includegraphics[width=0.1\textwidth, height=0.1\textwidth]{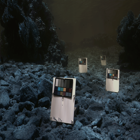} &
        \includegraphics[width=0.1\textwidth, height=0.1\textwidth]{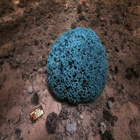} &
        \includegraphics[width=0.1\textwidth, height=0.1\textwidth]{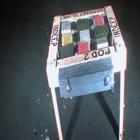} &
        \includegraphics[width=0.1\textwidth, height=0.1\textwidth]{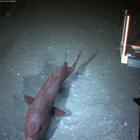} &
        \includegraphics[width=0.1\textwidth, height=0.1\textwidth]{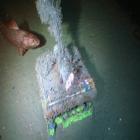} &
        \includegraphics[width=0.1\textwidth, height=0.1\textwidth]{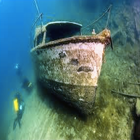} &
        \includegraphics[width=0.1\textwidth, height=0.1\textwidth]{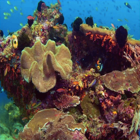} &
        \includegraphics[width=0.1\textwidth, height=0.1\textwidth]{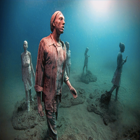} \\[-0.1em]

        \textbf{\scriptsize\textit{UDNet}} &
        \includegraphics[width=0.1\textwidth, height=0.1\textwidth]{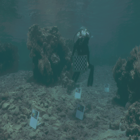} &
        \includegraphics[width=0.1\textwidth, height=0.1\textwidth]{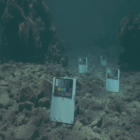} &
        \includegraphics[width=0.1\textwidth, height=0.1\textwidth]{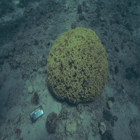} &
        \includegraphics[width=0.1\textwidth, height=0.1\textwidth]{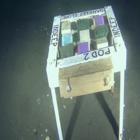} &
        \includegraphics[width=0.1\textwidth, height=0.1\textwidth]{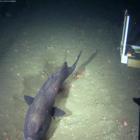} &
        \includegraphics[width=0.1\textwidth, height=0.1\textwidth]{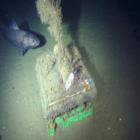} &
        \includegraphics[width=0.1\textwidth, height=0.1\textwidth]{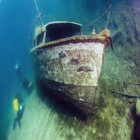} &
        \includegraphics[width=0.1\textwidth, height=0.1\textwidth]{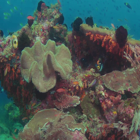} &
        \includegraphics[width=0.1\textwidth, height=0.1\textwidth]{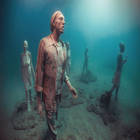} \\[-0.1em]

        \textbf{\scriptsize\textit{UST}} &
        \includegraphics[width=0.1\textwidth, height=0.1\textwidth]{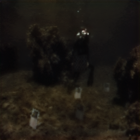} &
        \includegraphics[width=0.1\textwidth, height=0.1\textwidth]{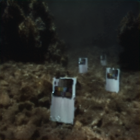} &
        \includegraphics[width=0.1\textwidth, height=0.1\textwidth]{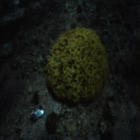} &
        \includegraphics[width=0.1\textwidth, height=0.1\textwidth]{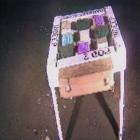} &
        \includegraphics[width=0.1\textwidth, height=0.1\textwidth]{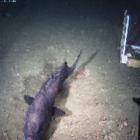} &
        \includegraphics[width=0.1\textwidth, height=0.1\textwidth]{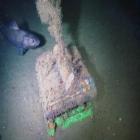} &
        \includegraphics[width=0.1\textwidth, height=0.1\textwidth]{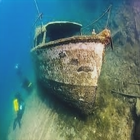} &
        \includegraphics[width=0.1\textwidth, height=0.1\textwidth]{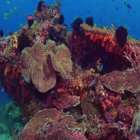} &
        \includegraphics[width=0.1\textwidth, height=0.1\textwidth]{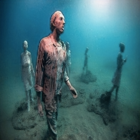} \\[-0.1em]

        \textbf{\scriptsize\textit{DIVER}} &
        \includegraphics[width=0.1\textwidth, height=0.1\textwidth]{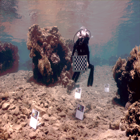} &
        \includegraphics[width=0.1\textwidth, height=0.1\textwidth]{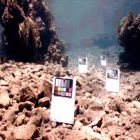} &
        \includegraphics[width=0.1\textwidth, height=0.1\textwidth]{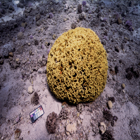} &
        \includegraphics[width=0.1\textwidth, height=0.1\textwidth]{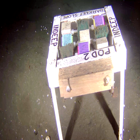} &
        \includegraphics[width=0.1\textwidth, height=0.1\textwidth]{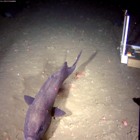} &
        \includegraphics[width=0.1\textwidth, height=0.1\textwidth]{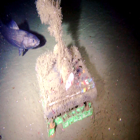} &
        \includegraphics[width=0.1\textwidth, height=0.1\textwidth]{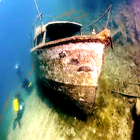} &
        \includegraphics[width=0.1\textwidth, height=0.1\textwidth]{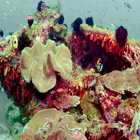} &
        \includegraphics[width=0.1\textwidth, height=0.1\textwidth]{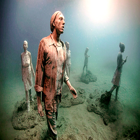} \\[-0.1em]
        
    \end{tabular}
    \caption{Visual comparison of underwater image enhancement results on three datasets: \textit{SeaThru}~\cite{akkaynak2019}, \textit{OceanDark}~\cite{marques2020}, and \textit{USOD10K}~\cite{Hong2025}. Each row corresponds to a SOTA methods: traditional approaches (IBLA~\cite{peng2017}, DCP~\cite{he2010}, UDCP~\cite{drews2013}, and ULAP~\cite{song2018}), deep learning-based methods (P2CNet~\cite{rao2023}, Phaseformer (PF)~\cite{khan2025}, WaterNet~\cite{li2019}, UDNet~\cite{saleh2025adaptive}, and the U-Shape Transformer (UST)~\cite{peng2023}), and our proposed method DIVER.}

    \label{fig:comparison_all_1}
\end{figure*}

\subsection{Baseline Compared}

We benchmark DIVER against eleven representative underwater enhancement methods, covering both traditional priors and modern deep learning approaches. The traditional baselines include IBLA~\cite{peng2017}, DCP~\cite{he2010}, UDCP~\cite{drews2013}, and ULAP~\cite{song2018}, capture common restoration assumptions but struggle under strong spectral shifts, non-uniform illumination, and high turbidity. We additionally benchmark DIVER against six learning-based approaches: P2CNet~\cite{rao2023}, Phaseformer~\cite{khan2025}, WaterNet~\cite{li2019}, UDNet~\cite{saleh2025adaptive}, Spectroformer, and the U-Shape Transformer~\cite{peng2023}. These methods achieve strong results within specific training domains but often exhibit domain sensitivity due to dataset-dependent color and illumination statistics. In contrast, DIVER integrates empirical correction, adaptive learning, and physics-guided modeling, enabling consistent restoration of spectral balance, contrast, and radiance across diverse underwater conditions.

\begin{figure*}[htbp]
    \centering

    \setlength{\tabcolsep}{0pt}
    \renewcommand{\arraystretch}{0}

    \begin{tabular}{@{}l
        @{\hspace{0.3em}}*{3}{@{}c@{}} 
        @{\hspace{1em}}                
        *{3}{@{}c@{}}                  
        @{\hspace{1em}}                
        *{3}{@{}c@{}}@{}               
        @{}}
        
        & \multicolumn{3}{c}{\textbf{(a)~\textit{U45}}} &
          \multicolumn{3}{c}{\textbf{(b)~\textit{FISHTRAC}}} &
          \multicolumn{3}{c}{\textbf{(c)~\textit{UIEB}}} \\[0.3em]  
        \textbf{\scriptsize\textit{Raw}} &
        \includegraphics[width=0.1\textwidth,height=0.1\textwidth]{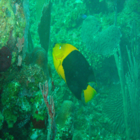} &
        \includegraphics[width=0.1\textwidth,height=0.1\textwidth]{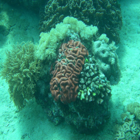} &
        \includegraphics[width=0.1\textwidth, height=0.1\textwidth]{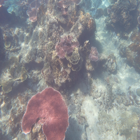} &
        \includegraphics[width=0.1\textwidth, height=0.1\textwidth]{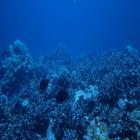} &
        \includegraphics[width=0.1\textwidth, height=0.1\textwidth]{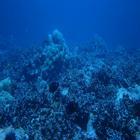} &
        \includegraphics[width=0.1\textwidth, height=0.1\textwidth]{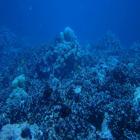} &
        \includegraphics[width=0.1\textwidth, height=0.1\textwidth]{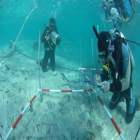} &
        \includegraphics[width=0.1\textwidth, height=0.1\textwidth]{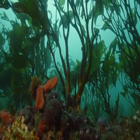} &
        \includegraphics[width=0.1\textwidth, height=0.1\textwidth]{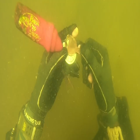} \\[-0.1em]
        
        \textbf{\scriptsize\textit{IBLA}} &
        \includegraphics[width=0.1\textwidth,height=0.1\textwidth]{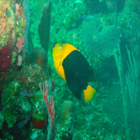} &
        \includegraphics[width=0.1\textwidth,height=0.1\textwidth]{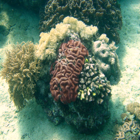} &
        \includegraphics[width=0.1\textwidth, height=0.1\textwidth]{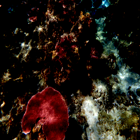} &
        \includegraphics[width=0.1\textwidth, height=0.1\textwidth]{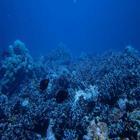} &
        \includegraphics[width=0.1\textwidth, height=0.1\textwidth]{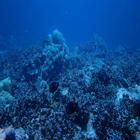} &
        \includegraphics[width=0.1\textwidth, height=0.1\textwidth]{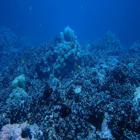} &
        \includegraphics[width=0.1\textwidth, height=0.1\textwidth]{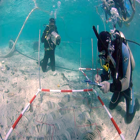} &
        \includegraphics[width=0.1\textwidth, height=0.1\textwidth]{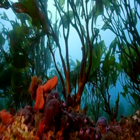} &
        \includegraphics[width=0.1\textwidth, height=0.1\textwidth]{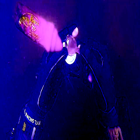} \\[-0.1em]

        \textbf{\scriptsize\textit{DCP}} &
        \includegraphics[width=0.1\textwidth, height=0.1\textwidth]{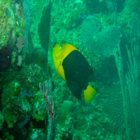} &
        \includegraphics[width=0.1\textwidth, height=0.1\textwidth]{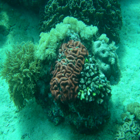} &
        \includegraphics[width=0.1\textwidth, height=0.1\textwidth]{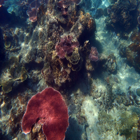} &
        \includegraphics[width=0.1\textwidth, height=0.1\textwidth]{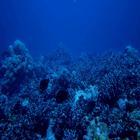} &
        \includegraphics[width=0.1\textwidth, height=0.1\textwidth]{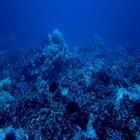} &
        \includegraphics[width=0.1\textwidth, height=0.1\textwidth]{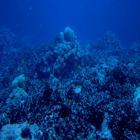} &
        \includegraphics[width=0.1\textwidth, height=0.1\textwidth]{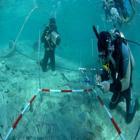} &
        \includegraphics[width=0.1\textwidth, height=0.1\textwidth]{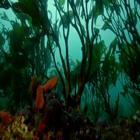} &
        \includegraphics[width=0.1\textwidth, height=0.1\textwidth]{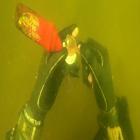} \\[-0.1em]

        \textbf{\scriptsize\textit{UDCP}} &
        \includegraphics[width=0.1\textwidth, height=0.1\textwidth]{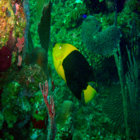} &
        \includegraphics[width=0.1\textwidth, height=0.1\textwidth]{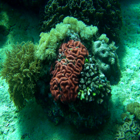} &
        \includegraphics[width=0.1\textwidth, height=0.1\textwidth]{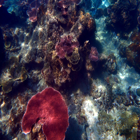} &
        \includegraphics[width=0.1\textwidth, height=0.1\textwidth]{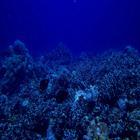} &
        \includegraphics[width=0.1\textwidth, height=0.1\textwidth]{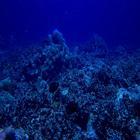} &
        \includegraphics[width=0.1\textwidth, height=0.1\textwidth]{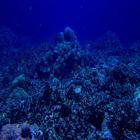} &
        \includegraphics[width=0.1\textwidth, height=0.1\textwidth]{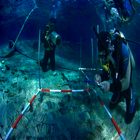} &
        \includegraphics[width=0.1\textwidth, height=0.1\textwidth]{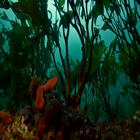} &
        \includegraphics[width=0.1\textwidth, height=0.1\textwidth]{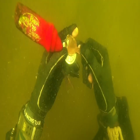} \\[-0.1em]

        \textbf{\scriptsize\textit{ULAP}} &
        \includegraphics[width=0.1\textwidth, height=0.1\textwidth]{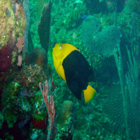} &
        \includegraphics[width=0.1\textwidth, height=0.1\textwidth]{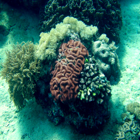} &
        \includegraphics[width=0.1\textwidth, height=0.1\textwidth]{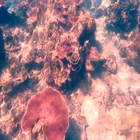} &
        \includegraphics[width=0.1\textwidth, height=0.1\textwidth]{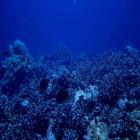} &
        \includegraphics[width=0.1\textwidth, height=0.1\textwidth]{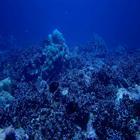} &
        \includegraphics[width=0.1\textwidth, height=0.1\textwidth]{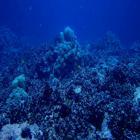} &
        \includegraphics[width=0.1\textwidth, height=0.1\textwidth]{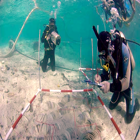} &
        \includegraphics[width=0.1\textwidth, height=0.1\textwidth]{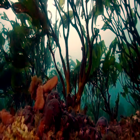} &
        \includegraphics[width=0.1\textwidth, height=0.1\textwidth]{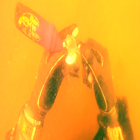} \\[-0.1em]

        \textbf{\scriptsize\textit{P2CNet}} &
        \includegraphics[width=0.1\textwidth, height=0.1\textwidth]{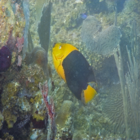} &
        \includegraphics[width=0.1\textwidth, height=0.1\textwidth]{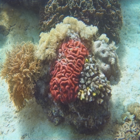} &
        \includegraphics[width=0.1\textwidth, height=0.1\textwidth]{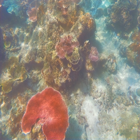} &
        \includegraphics[width=0.1\textwidth, height=0.1\textwidth]{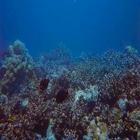} &
        \includegraphics[width=0.1\textwidth, height=0.1\textwidth]{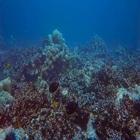} &
        \includegraphics[width=0.1\textwidth, height=0.1\textwidth]{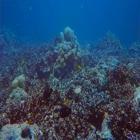} &
        \includegraphics[width=0.1\textwidth, height=0.1\textwidth]{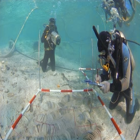} &
        \includegraphics[width=0.1\textwidth, height=0.1\textwidth]{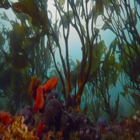} &
        \includegraphics[width=0.1\textwidth, height=0.1\textwidth]{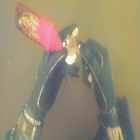} \\[-0.1em]

        \textbf{\scriptsize\textit{PF}} &
        \includegraphics[width=0.1\textwidth, height=0.1\textwidth]{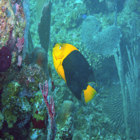} &
        \includegraphics[width=0.1\textwidth, height=0.1\textwidth]{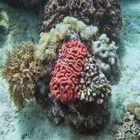} &
        \includegraphics[width=0.1\textwidth, height=0.1\textwidth]{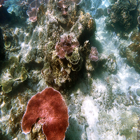} &
        \includegraphics[width=0.1\textwidth, height=0.1\textwidth]{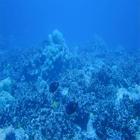} &
        \includegraphics[width=0.1\textwidth, height=0.1\textwidth]{JournalFigures/Comparisions/FishTrac/Phaseformer/img00115.jpg} &
        \includegraphics[width=0.1\textwidth, height=0.1\textwidth]{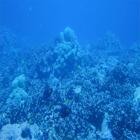} &
        \includegraphics[width=0.1\textwidth, height=0.1\textwidth]{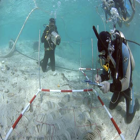} &
        \includegraphics[width=0.1\textwidth, height=0.1\textwidth]{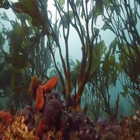} &
        \includegraphics[width=0.1\textwidth, height=0.1\textwidth]{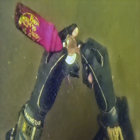} \\[-0.1em]

        \textbf{\scriptsize\textit{WaterNet}} &
        \includegraphics[width=0.1\textwidth, height=0.1\textwidth]{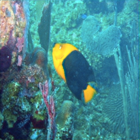} &
        \includegraphics[width=0.1\textwidth, height=0.1\textwidth]{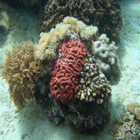} &
        \includegraphics[width=0.1\textwidth, height=0.1\textwidth]{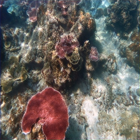} &
        \includegraphics[width=0.1\textwidth, height=0.1\textwidth]{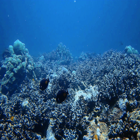} &
        \includegraphics[width=0.1\textwidth, height=0.1\textwidth]{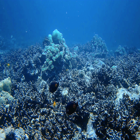} &
        \includegraphics[width=0.1\textwidth, height=0.1\textwidth]{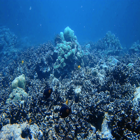} &
        \includegraphics[width=0.1\textwidth, height=0.1\textwidth]{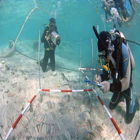} &
        \includegraphics[width=0.1\textwidth, height=0.1\textwidth]{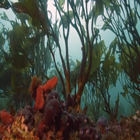} &
        \includegraphics[width=0.1\textwidth, height=0.1\textwidth]{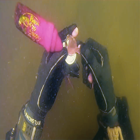} \\[-0.1em]

        \textbf{\scriptsize\textit{UDNet}} &
        \includegraphics[width=0.1\textwidth, height=0.1\textwidth]{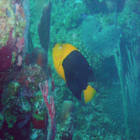} &
        \includegraphics[width=0.1\textwidth, height=0.1\textwidth]{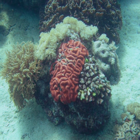} &
        \includegraphics[width=0.1\textwidth, height=0.1\textwidth]{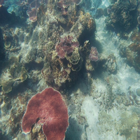} &
        \includegraphics[width=0.1\textwidth, height=0.1\textwidth]{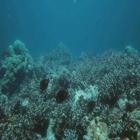} &
        \includegraphics[width=0.1\textwidth, height=0.1\textwidth]{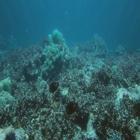} &
        \includegraphics[width=0.1\textwidth, height=0.1\textwidth]{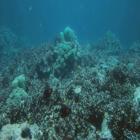} &
        \includegraphics[width=0.1\textwidth, height=0.1\textwidth]{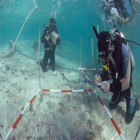} &
        \includegraphics[width=0.1\textwidth, height=0.1\textwidth]{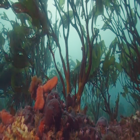} &
        \includegraphics[width=0.1\textwidth, height=0.1\textwidth]{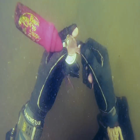} \\[-0.1em]

        \textbf{\scriptsize\textit{UST}} &
        \includegraphics[width=0.1\textwidth, height=0.1\textwidth]{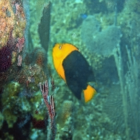} &
        \includegraphics[width=0.1\textwidth, height=0.1\textwidth]{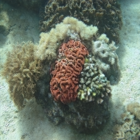} &
        \includegraphics[width=0.1\textwidth, height=0.1\textwidth]{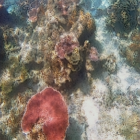} &
        \includegraphics[width=0.1\textwidth, height=0.1\textwidth]{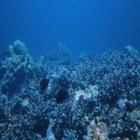} &
        \includegraphics[width=0.1\textwidth, height=0.1\textwidth]{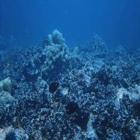} &
        \includegraphics[width=0.1\textwidth, height=0.1\textwidth]{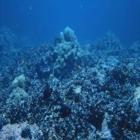} &
        \includegraphics[width=0.1\textwidth, height=0.1\textwidth]{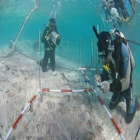} &
        \includegraphics[width=0.1\textwidth, height=0.1\textwidth]{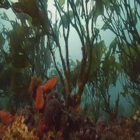} &
        \includegraphics[width=0.1\textwidth, height=0.1\textwidth]{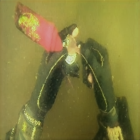} \\[-0.1em]

        \textbf{\scriptsize\textit{DIVER}} &
        \includegraphics[width=0.1\textwidth, height=0.1\textwidth]{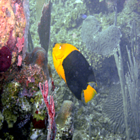} &
        \includegraphics[width=0.1\textwidth, height=0.1\textwidth]{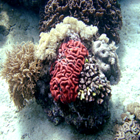} &
        \includegraphics[width=0.1\textwidth, height=0.1\textwidth]{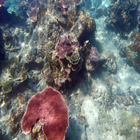} &
        \includegraphics[width=0.1\textwidth, height=0.1\textwidth]{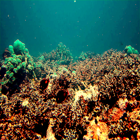} &
        \includegraphics[width=0.1\textwidth, height=0.1\textwidth]{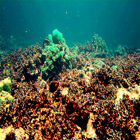} &
        \includegraphics[width=0.1\textwidth, height=0.1\textwidth]{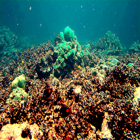} &
        \includegraphics[width=0.1\textwidth, height=0.1\textwidth]{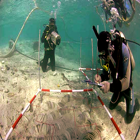} &
        \includegraphics[width=0.1\textwidth, height=0.1\textwidth]{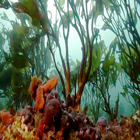} &
        \includegraphics[width=0.1\textwidth, height=0.1\textwidth]{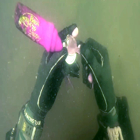} \\[-0.1em]
        
    \end{tabular}
    \caption{Visual comparison of underwater image enhancement results on three datasets: \textit{U45}~\cite{drews2013}, \textit{FISHTRAC}~\cite{dawkins2024}, and \textit{UIEB}~\cite{li2019}. Each row corresponds to a SOTA methods: traditional approaches (IBLA~\cite{peng2017}, DCP~\cite{he2010}, UDCP~\cite{drews2013}, and ULAP~\cite{song2018}), deep learning-based methods (P2CNet~\cite{rao2023}, Phaseformer (PF)~\cite{khan2025}, WaterNet~\cite{li2019}, UDNet~\cite{saleh2025adaptive}, and the U-Shape Transformer (UST)~\cite{peng2023}), and our proposed method DIVER.}

    \label{fig:comparison_all_2}
\end{figure*}


\begin{figure*}[htbp]
    \centering

    \setlength{\tabcolsep}{0pt}
    \renewcommand{\arraystretch}{0}

    \begin{tabular}{@{}l
        @{\hspace{0.3em}}*{3}{@{}c@{}} 
        @{\hspace{1em}}                
        *{3}{@{}c@{}}                  
        @{\hspace{1em}}                
        *{3}{@{}c@{}}@{}               
        @{}}
        
        & \multicolumn{3}{c}{\textbf{(a)~\textit{UFO-120}}} &
          \multicolumn{3}{c}{\textbf{(b)~\textit{EUVP}}} &
          \multicolumn{3}{c}{\textbf{(c)~\textit{LSUI}}} \\[0.3em]  
        \textbf{\scriptsize\textit{Raw}} &
        \includegraphics[width=0.1\textwidth,height=0.1\textwidth]{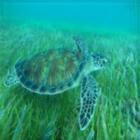} &
        \includegraphics[width=0.1\textwidth,height=0.1\textwidth]{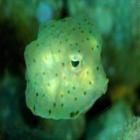} &
        \includegraphics[width=0.1\textwidth, height=0.1\textwidth]{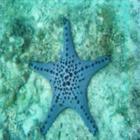} &
        \includegraphics[width=0.1\textwidth, height=0.1\textwidth]{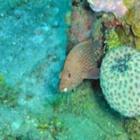} &
        \includegraphics[width=0.1\textwidth, height=0.1\textwidth]{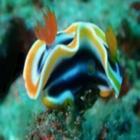} &
        \includegraphics[width=0.1\textwidth, height=0.1\textwidth]{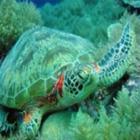} &
        \includegraphics[width=0.1\textwidth, height=0.1\textwidth]{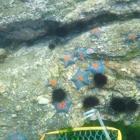} &
        \includegraphics[width=0.1\textwidth, height=0.1\textwidth]{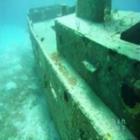} &
        \includegraphics[width=0.1\textwidth, height=0.1\textwidth]{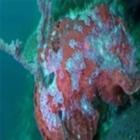} \\[-0.1em]
        
        \textbf{\scriptsize\textit{IBLA}} &
        \includegraphics[width=0.1\textwidth,height=0.1\textwidth]{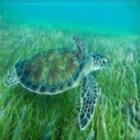} &
        \includegraphics[width=0.1\textwidth,height=0.1\textwidth]{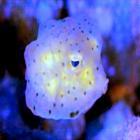} &
        \includegraphics[width=0.1\textwidth, height=0.1\textwidth]{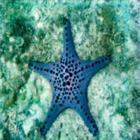} &
        \includegraphics[width=0.1\textwidth, height=0.1\textwidth]{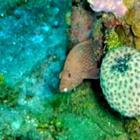} &
        \includegraphics[width=0.1\textwidth, height=0.1\textwidth]{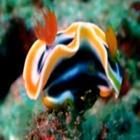} &
        \includegraphics[width=0.1\textwidth, height=0.1\textwidth]{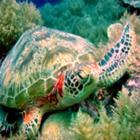} &
        \includegraphics[width=0.1\textwidth, height=0.1\textwidth]{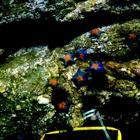} &
        \includegraphics[width=0.1\textwidth, height=0.1\textwidth]{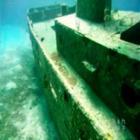} &
        \includegraphics[width=0.1\textwidth, height=0.1\textwidth]{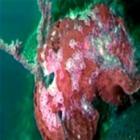} \\[-0.1em]

        \textbf{\scriptsize\textit{DCP}} &
        \includegraphics[width=0.1\textwidth, height=0.1\textwidth]{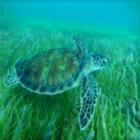} &
        \includegraphics[width=0.1\textwidth, height=0.1\textwidth]{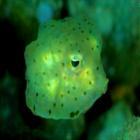} &
        \includegraphics[width=0.1\textwidth, height=0.1\textwidth]{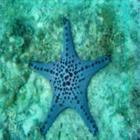} &
        \includegraphics[width=0.1\textwidth, height=0.1\textwidth]{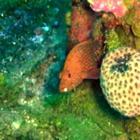} &
        \includegraphics[width=0.1\textwidth, height=0.1\textwidth]{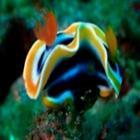} &
        \includegraphics[width=0.1\textwidth, height=0.1\textwidth]{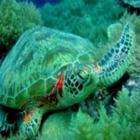} &
        \includegraphics[width=0.1\textwidth, height=0.1\textwidth]{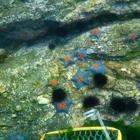} &
        \includegraphics[width=0.1\textwidth, height=0.1\textwidth]{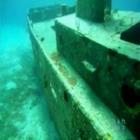} &
        \includegraphics[width=0.1\textwidth, height=0.1\textwidth]{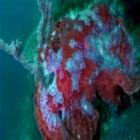} \\[-0.1em]

        \textbf{\scriptsize\textit{UDCP}} &
        \includegraphics[width=0.1\textwidth, height=0.1\textwidth]{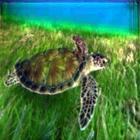} &
        \includegraphics[width=0.1\textwidth, height=0.1\textwidth]{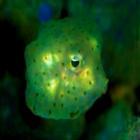} &
        \includegraphics[width=0.1\textwidth, height=0.1\textwidth]{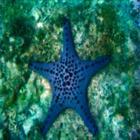} &
        \includegraphics[width=0.1\textwidth, height=0.1\textwidth]{JournalFigures/Comparisions/EUVP/UDCP/test_p117__UDCP.jpg} &
        \includegraphics[width=0.1\textwidth, height=0.1\textwidth]{JournalFigures/Comparisions/EUVP/UDCP/test_p286__UDCP.jpg} &
        \includegraphics[width=0.1\textwidth, height=0.1\textwidth]{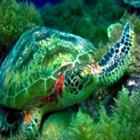} &
        \includegraphics[width=0.1\textwidth, height=0.1\textwidth]{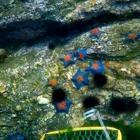} &
        \includegraphics[width=0.1\textwidth, height=0.1\textwidth]{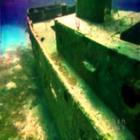} &
        \includegraphics[width=0.1\textwidth, height=0.1\textwidth]{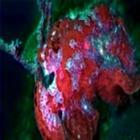} \\[-0.1em]

        \textbf{\scriptsize\textit{ULAP}} &
        \includegraphics[width=0.1\textwidth, height=0.1\textwidth]{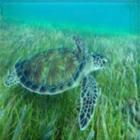} &
        \includegraphics[width=0.1\textwidth, height=0.1\textwidth]{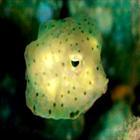} &
        \includegraphics[width=0.1\textwidth, height=0.1\textwidth]{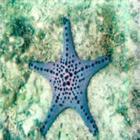} &
        \includegraphics[width=0.1\textwidth, height=0.1\textwidth]{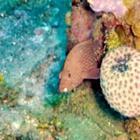} &
        \includegraphics[width=0.1\textwidth, height=0.1\textwidth]{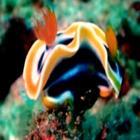} &
        \includegraphics[width=0.1\textwidth, height=0.1\textwidth]{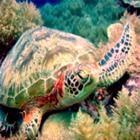} &
        \includegraphics[width=0.1\textwidth, height=0.1\textwidth]{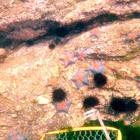} &
        \includegraphics[width=0.1\textwidth, height=0.1\textwidth]{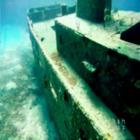} &
        \includegraphics[width=0.1\textwidth, height=0.1\textwidth]{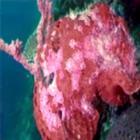} \\[-0.1em]

        \textbf{\scriptsize\textit{P2CNet}} &
        \includegraphics[width=0.1\textwidth, height=0.1\textwidth]{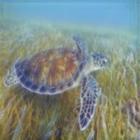} &
        \includegraphics[width=0.1\textwidth, height=0.1\textwidth]{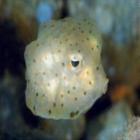} &
        \includegraphics[width=0.1\textwidth, height=0.1\textwidth]{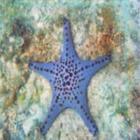} &
        \includegraphics[width=0.1\textwidth, height=0.1\textwidth]{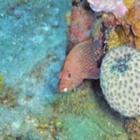} &
        \includegraphics[width=0.1\textwidth, height=0.1\textwidth]{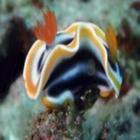} &
        \includegraphics[width=0.1\textwidth, height=0.1\textwidth]{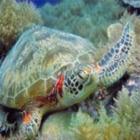} &
        \includegraphics[width=0.1\textwidth, height=0.1\textwidth]{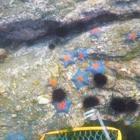} &
        \includegraphics[width=0.1\textwidth, height=0.1\textwidth]{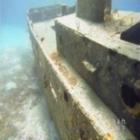} &
        \includegraphics[width=0.1\textwidth, height=0.1\textwidth]{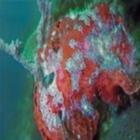} \\[-0.1em]

        \textbf{\scriptsize\textit{PF}} &
        \includegraphics[width=0.1\textwidth, height=0.1\textwidth]{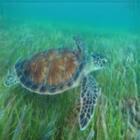} &
        \includegraphics[width=0.1\textwidth, height=0.1\textwidth]{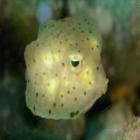} &
        \includegraphics[width=0.1\textwidth, height=0.1\textwidth]{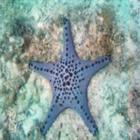} &
        \includegraphics[width=0.1\textwidth, height=0.1\textwidth]{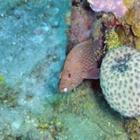} &
        \includegraphics[width=0.1\textwidth, height=0.1\textwidth]{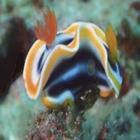} &
        \includegraphics[width=0.1\textwidth, height=0.1\textwidth]{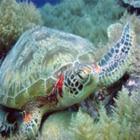} &
        \includegraphics[width=0.1\textwidth, height=0.1\textwidth]{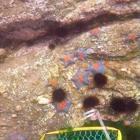} &
        \includegraphics[width=0.1\textwidth, height=0.1\textwidth]{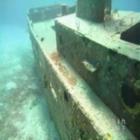} &
        \includegraphics[width=0.1\textwidth, height=0.1\textwidth]{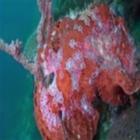} \\[-0.1em]

        \textbf{\scriptsize\textit{WaterNet}} &
        \includegraphics[width=0.1\textwidth, height=0.1\textwidth]{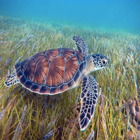} &
        \includegraphics[width=0.1\textwidth, height=0.1\textwidth]{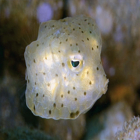} &
        \includegraphics[width=0.1\textwidth, height=0.1\textwidth]{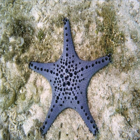} &
        \includegraphics[width=0.1\textwidth, height=0.1\textwidth]{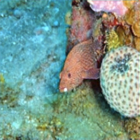} &
        \includegraphics[width=0.1\textwidth, height=0.1\textwidth]{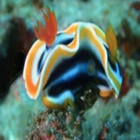} &
        \includegraphics[width=0.1\textwidth, height=0.1\textwidth]{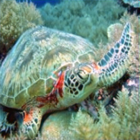} &
        \includegraphics[width=0.1\textwidth, height=0.1\textwidth]{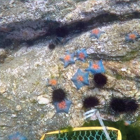} &
        \includegraphics[width=0.1\textwidth, height=0.1\textwidth]{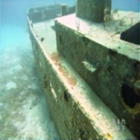} &
        \includegraphics[width=0.1\textwidth, height=0.1\textwidth]{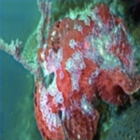} \\[-0.1em]

        \textbf{\scriptsize\textit{UDNet}} &
        \includegraphics[width=0.1\textwidth, height=0.1\textwidth]{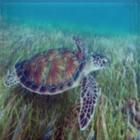} &
        \includegraphics[width=0.1\textwidth, height=0.1\textwidth]{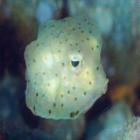} &
        \includegraphics[width=0.1\textwidth, height=0.1\textwidth]{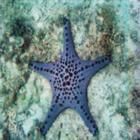} &
        \includegraphics[width=0.1\textwidth, height=0.1\textwidth]{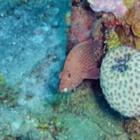} &
        \includegraphics[width=0.1\textwidth, height=0.1\textwidth]{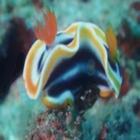} &
        \includegraphics[width=0.1\textwidth, height=0.1\textwidth]{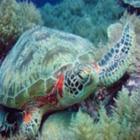} &
        \includegraphics[width=0.1\textwidth, height=0.1\textwidth]{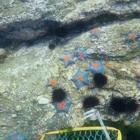} &
        \includegraphics[width=0.1\textwidth, height=0.1\textwidth]{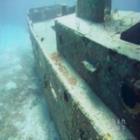} &
        \includegraphics[width=0.1\textwidth, height=0.1\textwidth]{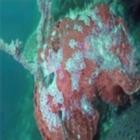} \\[-0.1em]

        \textbf{\scriptsize\textit{UST}} &
        \includegraphics[width=0.1\textwidth, height=0.1\textwidth]{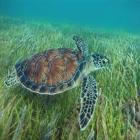} &
        \includegraphics[width=0.1\textwidth, height=0.1\textwidth]{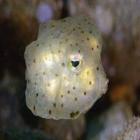} &
        \includegraphics[width=0.1\textwidth, height=0.1\textwidth]{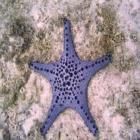} &
        \includegraphics[width=0.1\textwidth, height=0.1\textwidth]{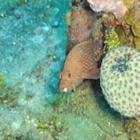} &
        \includegraphics[width=0.1\textwidth, height=0.1\textwidth]{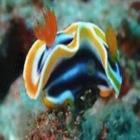} &
        \includegraphics[width=0.1\textwidth, height=0.1\textwidth]{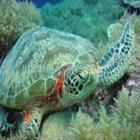} &
        \includegraphics[width=0.1\textwidth, height=0.1\textwidth]{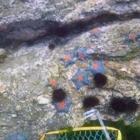} &
        \includegraphics[width=0.1\textwidth, height=0.1\textwidth]{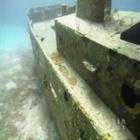} &
        \includegraphics[width=0.1\textwidth, height=0.1\textwidth]{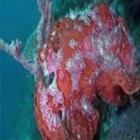} \\[-0.1em]

        \textbf{\scriptsize\textit{DIVER}} &
        \includegraphics[width=0.1\textwidth, height=0.1\textwidth]{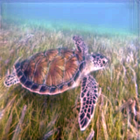} &
        \includegraphics[width=0.1\textwidth, height=0.1\textwidth]{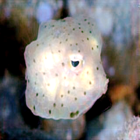} &
        \includegraphics[width=0.1\textwidth, height=0.1\textwidth]{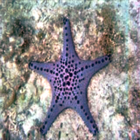} &
        \includegraphics[width=0.1\textwidth, height=0.1\textwidth]{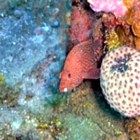} &
        \includegraphics[width=0.1\textwidth, height=0.1\textwidth]{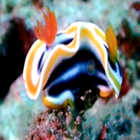} &
        \includegraphics[width=0.1\textwidth, height=0.1\textwidth]{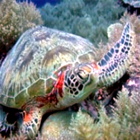} &
        \includegraphics[width=0.1\textwidth, height=0.1\textwidth]{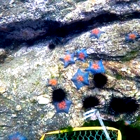} &
        \includegraphics[width=0.1\textwidth, height=0.1\textwidth]{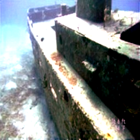} &
        \includegraphics[width=0.1\textwidth, height=0.1\textwidth]{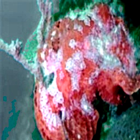} \\[-0.1em]
     \end{tabular}
    \caption{Visual comparison of underwater image enhancement results on three datasets: \textit{UFO-120}~\cite{wang2019deep}, \textit{EUVP}~\cite{islam2020}, and \textit{LSUI}~\cite{peng2023}. Each row corresponds to a SOTA methods: traditional approaches (IBLA~\cite{peng2017}, DCP~\cite{he2010}, UDCP~\cite{drews2013}, and ULAP~\cite{song2018}), deep learning-based methods (P2CNet~\cite{rao2023}, Phaseformer (PF)~\cite{khan2025}, WaterNet~\cite{li2019}, UDNet~\cite{saleh2025adaptive}, and the U-Shape Transformer (UST)~\cite{peng2023}), and our proposed method DIVER.}

    \label{fig:comparison_all_3}
\end{figure*}

\subsection{DIVER Implementation Details}
The proposed architecture was trained on a variety of datasets which consists of images with varying resolutions as it’s input. The setup of a Linux host consisted of an NVIDIA GeForce RTX 4090 GPU was used for training the models. The training of each of the network’s component was carried out with a learning rate of $0.001$ with ADAM optimizer. For IlluminateNet training, a batch size of 8 was taken with 150 iterations, while for Hydro-OpticNet transformation, a batch size of 10 with 50 iterations was considered.

The DIVER architecture was implemented using PyTorch and OpenCV frameworks, and early stopping with a patience of 20 epochs was employed, with the checkpoint with the lowest overall loss was retained to minimize overfitting. For each module, computational training time is as follows: the IlluminateNet takes on an average about 10 minutes to train, and the Hydro-OpticNet on an average takes 20 minutes to train. So, the entire DIVER model takes 30 - 35 minutes to train along with AOCM. The computational training time also depends on the dataset size. As mentioned in section~\ref{sec:diverlearning}, the loss functions used for training our architecture comprise a combination of Grayworld Loss (\(\mathcal{L}_{G}\)), Luminous Loss (\(\mathcal{L}_{L}\)), Adaptive Huber Loss (\(\mathcal{L}_{H}\)), Color Consistency Loss (\(\mathcal{L}_{C}\)), and Sobel Edge Loss (\(\mathcal{L}_{S}\)). As per computational complexity, DIVER takes 0.01sec per image on average as inference time, which is comparable to some of existing models in the same segment. However, it is important to note that the computational requirements may vary depending on the size and complexity of the input images, as well as the hardware used for training and inference.

\subsection{Qualitative Performance Comparison of DIVER}
Qualitative evaluation across multiple underwater datasets determines that the  proposed DIVER architecture produces perceptually better results with strong color fidelity and effective contrast restoration across diverse water types and lighting conditions. Figures~\ref{fig:comparison_all_1} - \ref{fig:comparison_all_3} illustrate visual comparisons of underwater image enhancement results across eight benchmark datasets See-Thru~\cite{akkaynak2019},OceanDark~\cite{marques2020}, USOD10K~\cite{islam2020},  U45~\cite{drews2013}, FISHTRAC \cite{dawkins2024}, UIEB~\cite{akkaynak2019}, UFO-120 \cite{dawkins2024}, LSUI~\cite{marques2020}, and EUVP~\cite{islam2020} highlighting the perceptual quality distinctions between the proposed DIVER architecture and SOTA methods.

As observed in Figure~\ref{fig:comparison_all_1} (a) illustrates qualitative comparisons on the SeaThru dataset. DIVER effectively removes the greenish cast while significantly improving illumination and restoring scene visibility. Phaseformer retains structural details better than IBLA, DCP, UDCP, ULAP, and P2CNet, which tend to produce overly dark images. WaterNet fails to correct color imbalance or recover illumination while UDNet improves brightness to some extent but leaves haze residuals. Meanwhile, UST struggles with low-light underwater scenes. OceanDark dataset has extremely low-light and localized artificial lighting as shown in Figure~\ref{fig:comparison_all_1} (b). DIVER successfully restores balanced illumination and realistic color appearance under artificial lighting conditions. IBLA introduces an excessive purple shift, and Phaseformer yields unnatural brown tones, while WaterNet leans toward a bluish cast that suppresses structural contrast. Although UST recovers some color, its results remain visually flat with insufficient detail enhancement. By comparison, DIVER provides a clearer and more naturally illuminated representation. The USOD10K dataset has highly turbid scenes. As seen in Figure~\ref{fig:comparison_all_1} (c), classical methods (IBLA, DCP, UDCP) fail to remove dominant color distortions, and methods like P2CNet, Phaseformer, WaterNet, UDNet often amplify noise or produce oversaturated hues. DIVER maintains contrast and detail across all three sample scenes, demonstrating strong generalization to turbidity variations.

\begin{table*}
\centering

\caption{Comparison against published works on five \textbf{UNPAIRED datasets}, SeaThru~\cite{akkaynak2019}, OceanDark~\cite{marques2020}, USOD10K~\cite{islam2020}, FISHTRAC \cite{dawkins2024}, U45~\cite{drews2013}. Underwater image enhancement performance metric in terms of average UIQM, UCIQE values are shown, where ($ \uparrow $
) means higher is better, and ($ \downarrow $
) means lower is better. We represent the first best scores in \textcolor{blue}{BLUE} and second best score in \textcolor{orange}{ORANGE}.}
\renewcommand{\arraystretch}{1.5} 
\setlength{\tabcolsep}{6pt} 
\LARGE
\resizebox{\textwidth}{!}{%

\begin{tabular}{@{}ccccccccccccccccc@{}}

\hline
\textbf{Method}                                                        & \multicolumn{3}{c}{\textbf{SeaThru}}                     & \multicolumn{3}{c}{\textbf{OceanDark}}                   & \multicolumn{3}{c}{\textbf{USOD10K}}                     & \multicolumn{3}{c}{\textbf{FISHTRAC}}                    & \multicolumn{3}{c}{\textbf{U45}}                         \\ \hline
\textbf{}                                                              & \textbf{UCIQE ↑} & \textbf{UIQM ↑} & \textbf{BRISQUE  ↓} & \textbf{UCIQE ↑} & \textbf{UIQM ↑} & \textbf{BRISQUE  ↓} & \textbf{UCIQE ↑} & \textbf{UIQM ↑} & \textbf{BRISQUE  ↓} & \textbf{UCIQE ↑} & \textbf{UIQM ↑} & \textbf{BRISQUE  ↓} & \textbf{UCIQE ↑} & \textbf{UIQM ↑} & \textbf{BRISQUE  ↓} \\ \hline
\textbf{IBLA}                                                          & 0.294           & 1.327          & 29.698              & 0.800           & 1.571          & 37.121              & 0.786           & 1.480          & 37.319              & 0.538           & 1.364          & 34.161              & 0.685           & 1.708           & 23.991              \\ \hline
\textbf{DCP}                                                           & 0.252           & 0.427          & 40.808              & 0.676           & 1.728          & 33.329              & 0.683           & 1.986          & 34.027              & 0.621         & 1.316        & 34.636              & 0.604           & 2.257          & 21.478              \\ \hline
\textbf{UDCP}                                                          & 0.214          & 0.424          & 40.469              & 0.704           & 1.251          & 34.619              & 0.791           & 1.819          & 33.111              & 0.837         & 1.321        & 33.163              & 0.747           & 2.110          & 18.802              \\ \hline
\textbf{ULAP}                                                          & 0.168           & 0.471          & 39.578              & 0.767           & 1.936          & 33.665              & 0.773           & 1.922          & 33.411              & 0.608         & 1.402        & 31.598             & 0.640           & 2.301          & 19.056              \\ \hline
\textbf{UST} & 0.331           & 2.788          & 31.456              & 1.009           & \textbf{\textcolor{orange}{2.882}}          & 28.010              & 0.809           & \textbf{\textcolor{orange}{2.740}}          & \textbf{\textcolor{blue}{29.733}}              & 0.754           & 1.798          & 33.845              & 0.745           & \textbf{\textcolor{blue}{3.155}}          & 21.586              \\ \hline
\textbf{P2CNet}                                                        & 0.300           & 1.214          & \textbf{\textcolor{orange}{21.712}}              & \textbf{\textcolor{orange}{1.515}}           & \textbf{\textcolor{blue}{2.764}}          & \textbf{\textcolor{blue}{19.921}}              & \textbf{\textcolor{orange}{0.943}}           & 2.565          & \textbf{\textcolor{orange}{31.486}}              & 0.529          & 2.311          & 28.400              & 0.655           & 3.089          & \textbf{\textcolor{orange}{17.908}}              \\ \hline
\textbf{Phaseformer}                                                   & 0.763           & 1.230          & 39.552              & 1.056           & 1.933          & 30.957              & 0.503           & 1.952          & 43.822              & 0.489           & 1.390          & 42.075              & 0.781           & 3.052          & 41.647              \\ \hline
\textbf{WaterNet}                                                      & \textbf{\textcolor{orange}{1.503}}           & \textbf{\textcolor{orange}{2.923}}          & 26.686              & 1.126           & 2.365          & 27.940              & 0.961           & 2.521          & 38.495              & 0.729           & 2.166          & \textbf{\textcolor{orange}{27.896}}              & 0.757           & \textbf{\textcolor{orange}{3.236}}          & 22.062              \\ \hline
\textbf{UDNet}                                                         & 0.624           & 2.869          & 26.129              & 0.933           & 2.121         & 30.954              & 0.683           & 2.458          & 39.622              & 0.677           & \textbf{\textcolor{orange}{2.786}}          & 34.466              & 0.681           & 3.107          & 20.763              \\ \hline
\textbf{DIVER (Ours)}                                                   & \textbf{\textcolor{blue}{1.654}}           & \textbf{\textcolor{blue}{2.985}}          & \textbf{\textcolor{blue}{20.946}}              &\textbf{\textcolor{blue} {1.676}}           & 2.681          & \textbf{\textcolor{orange}{23.558}}              & \textbf{\textcolor{blue}{1.180}}           &\textbf{\textcolor{blue}{2.750}}          & 32.297              & \textbf{\textcolor{blue}{1.007}}           & \textbf{\textcolor{blue}{2.813}}          & \textbf{\textcolor{blue}{27.67}}              & \textbf{\textcolor{blue}{1.028}}           & 2.655          & \textbf{\textcolor{blue}{17.604}}              
                                     \\ \bottomrule
\label{tab:Non-ref scores_1}
\end{tabular}%
}
\end{table*}

On the U45 dataset, the underwater scenes exhibit strong green dominance and haziness. As illustrated in Figure~\ref{fig:comparison_all_2} (a), conventional priors such as IBLA, DCP, UDCP, and ULAP fail to jointly correct the green overcast and contrast degradation: IBLA produces faded textures, while ULAP amplifies brightness excessively, leading to washed-out coral regions. Learning-based models like WaterNet, UDNet and UST restore brightness but still retain dominant green tint and structural blur. In contrast, DIVER recovers the natural chromatic distribution across coral (red color) and fish (specially yellow color) and achieves visually balanced color restoration. FISHTRAC imagery is captured in deeper waters with strong bluish casts. DIVER effectively restores the attenuated red tone while enhancing scene visibility and recovering coral textures with minimal artifacts as shown in Figure~\ref{fig:comparison_all_2} (b). In contrast, conventional methods such as IBLA, DCP, UDCP, and ULAP either darken the scene or retain an excessive blue dominance that obscures details. Although recent deep learning approaches like Phaseformer, WaterNet, and UDNet improve contrast to some extent, they fail to remove the bluish tint. As illustrated in the Figure~\ref{fig:comparison_all_2} (b), the UIEB dataset includes diverse lighting—harsh artificial illumination, and yellow-magenta shifts common in diver-held lamps. Conventional methods create either over-enhanced colors (IBLA turning scenes purplish) or dull corrections (UDCP/ULAP leaves haze and color inaccuracy). Transformer-based restoration (UST, Phaseformer) and WaterNet stabilizes structure but may overshoot luminance and induces unnecessary yellow hues, degrading natural color perception especially on diver suits and skin tones. DIVER achieves a balanced restoration and removes both haze and color speckles introduced by hard lighting. Further, it maintains accurate skin tones and consistent chromaticity (specially pink hue restoration) across objects.


\begin{table*}
\centering

\caption{Comparison against published works on four \textbf{PAIRED DATASETS}, UIEB~\cite{li2019}, LSUI~\cite{peng2023}, EUVP~\cite{islam2020}, UFO-120 \cite{wang2019deep} .Underwater image enhancement performance metric in terms of average UIQM, UCIQE, PSNR and SSIM  values are shown, where ($ \uparrow $
) means higher is better, and ($ \downarrow $
) means lower is better.  We represent the first best scores in \textcolor{blue}{BLUE} and second best score in \textcolor{orange}{ORANGE}.}
\renewcommand{\arraystretch}{1.5} 
\setlength{\tabcolsep}{6pt} 
\huge
\resizebox{\textwidth}{!}{%
\begin{tabular}{ccccccccccccccccc}
\hline
\textbf{Method}       & \multicolumn{4}{c}{\textbf{UIEB}}                                  & \multicolumn{4}{c}{\textbf{LSUI}}                                  & \multicolumn{4}{c}{\textbf{EUVP}}                                  & \multicolumn{4}{c}{\textbf{UFO-120}}                               \\ \hline
\textbf{}             & \textbf{PSNR} & \textbf{SSIM} & \textbf{UCIQE ↑} & \textbf{UIQM ↑} & \textbf{PSNR} & \textbf{SSIM} & \textbf{UCIQE ↑} & \textbf{UIQM ↑} & \textbf{PSNR} & \textbf{SSIM} & \textbf{UCIQE ↑} & \textbf{UIQM ↑} & \textbf{PSNR} & \textbf{SSIM} & \textbf{UCIQE ↑} & \textbf{UIQM ↑} \\ \hline
\textbf{IBLA}         & 18.172        & 0.763        & 0.679           & 1.502          & 8.151        & 0.893        & 0.963           & 2.296          & 10.089       & 0.812        & 0.724            & 2.196           & 9.169        & 0.619        & 0.756          & 2.051          \\ \hline
\textbf{DCP}          & 14.406        & 0.631        & 0.585           & 1.291          & 8.132        & 0.853        & 0.747          & 1.776          & 8.678        & 0.829        & 0.678           & 1.974           & 8.669        & 0.674        & 0.665           & 2.019          \\ \hline
\textbf{UDCP}         & 11.060        & 0.467        & 0.768           & 1.093          & 7.654        & 0.661        & 0.669           & 1.945          & 7.473        & 0.659        & 0.773           & 1.900           & 7.646        & 0.529        & 0.786           & 1.964          \\ \hline
\textbf{ULAP}         & 17.042        & 0.709        & 0.707           & 1.332          & 7.959        & 0.895        & 0.592           & 2.324          & 9.242        & 0.803        & 0.757          & 2.193          & 8.862        & 0.651        & 0.751           & 2.261          \\ \hline
\textbf{Spectoformer} & 21.522        & 0.817       & 0.677           & 2.493           & 11.628        & \textbf{\textcolor{orange}{0.911}}        & 0.678           & 2.812           & 9.296        & 0.842        & 0.707           & 2.812           & 9.126        & \textbf{\textcolor{orange}{0.696}}        & 0.699           & 2.794          \\ \hline
\textbf{P2CNet}       & 18.864        & 0.768        & \textbf{\textcolor{orange}{1.265}}            & 2.495           & 10.061        & 0.763        & 0.839           & 2.907          & 11.405       & 0.865        & 1.154            & 2.939           & 10.272       & 0.681        & \textbf{\textcolor{blue}{1.272}}            & 2.959           \\ \hline
\textbf{Phaseformer}  & 21.474        & 0.854        & 0.751            & \textbf{\textcolor{blue}{2.565}}           & 9.821        & 0.834        & 0.881            & 2.155           & 8.987        & 0.843        & 0.811          & 2.684           & 13.059       & \textbf{\textcolor{blue}{0.727}}        & \textbf{\textcolor{orange}{1.163}}            & 2.246           \\ \hline
\textbf{WaterNet}     & \textbf{\textcolor{orange}{23.172}}        & \textbf{\textcolor{orange}{0.863}}        & 1.021            & \textbf{\textcolor{orange}{2.515}}           & \textbf{\textcolor{blue}{24.187}}        & \textbf{\textcolor{blue}{0.932}}        & 0.817            & \textbf{\textcolor{blue}{3.099}}           & \textbf{\textcolor{blue}{19.917}}        & \textbf{\textcolor{orange}{0.867}}        & \textbf{\textcolor{orange}{0.976}}            & 2.939           & \textbf{\textcolor{blue}{18.628}}       & 0.653        & 1.029            & \textbf{\textcolor{blue}{3.008}}           \\ \hline
\textbf{UDNet}        & 17.641       & 0.784        & 0.715            & 2.398           & 9.104        & 0.892        & \textbf{\textcolor{orange}{0.913}}            & \textbf{\textcolor{orange}{2.957}}           & 9.203        & 0.845        & 0.935            & \textbf{\textcolor{blue}{3.059}}           & 8.927       & 0.671        & 0.721            & 2.896           \\ \hline
\textbf{DIVER (Ours)}  & \textbf{\textcolor{blue}{23.987}}        & \textbf{\textcolor{blue}{0.928}}        & \textbf{\textcolor{blue}{1.269}}           & 2.397          & \textbf{\textcolor{orange}{18.284}}        & 0.812        & \textbf{\textcolor{blue}{1.457}}           & 2.259          & \textbf{\textcolor{orange}{15.071}}        & \textbf{\textcolor{blue}{0.881}}        & \textbf{\textcolor{blue}{1.279}}            & \textbf{\textcolor{orange}{2.942}}           & \textbf{\textcolor{orange}{14.347}}       & 0.664       & 1.069           & \textbf{\textcolor{orange}{2.997}}          
\\ \hline
\label{Ref scores_1}
\end{tabular}%
}
\end{table*}

On UFO-120 dataset which contains images with moderate haze and color shift, DIVER restores natural textures and corrects the dominant green bias while preserving fine details on starfish and coral as shown in Figure~\ref{fig:comparison_all_3}(a). In comparison, IBLA, DCP, UDCP, and ULAP either desaturate textures or retain excessive green casts. Transformer-based methods (UST, Phaseformer) and CNN methods (WaterNet, UDNet) improve brightness and sharpness but often introduce unnatural hues. DIVER offers a balanced enhancement without artificial color shifts. Note that all the deep learning methods perform good restoration on the scene with the starfish. As seen for EUVP in Figure~\ref{fig:comparison_all_3} (b), traditional and deep learning methods have restored the color decently, but fails to remove the green tint across all the images. However, DIVER achieves clearer visibility and accurate color neutrality. On LSUI, which contains challenging low-light, colorcast underwater scenes, DIVER effectively brightens images while maintaining color fidelity as illustrated in Figure~\ref{fig:comparison_all_3} (c). While ULAP and Phaseformer restores the color but sometimes fails to remove the green tint and haze, while UDCP generates greenish-yellow color casts. Methods like P2CNet and WaterNet show improved brightness but sometimes in expense of natural color representation.

Across all eight benchmark datasets, classical priors such as IBLA, DCP, UDCP, and ULAP consistently struggle with domain shifts, either over-darkening scenes (DCP/UDCP on SeaThru, Fish), introducing excessive purple/blue tints (IBLA on UIEB, UFO-120), or producing over-brightened, washed-out textures (ULAP on U45, EUVP). Deep learning-based models like P2CNet and Phaseformer provide improved visibility but frequently suffer from color artifacts—green dominance on SeaThru, over-enhanced coral reds on U45, and harsh lighting inconsistencies on UIEB. WaterNet enhances contrast well yet often sacrifices fine-texture fidelity (SeaThru, EUVP, FISHTRAC), while UDNet and UST display improved structure preservation but retain unbalanced hues or hazy remnants across datasets (USOD10K, LSUI). In contrast, DIVER maintains consistent natural color tones on U45 (successful recovery of yellow fish and coral reds), corrects severe blue attenuation on FISHTRAC, and preserves illumination realism on UIEB without pink/yellow artifacts. On challenging low-light and turbid datasets like SeaThru, OceanDark, USOD10K, LSUI, DIVER reveals scene features lost by competing methods, demonstrating robust generalization to both spectral imbalance and illumination degradation. Overall, DIVER is the only method that consistently delivers enhanced contrast, accurate color compensation, and structural details across all dataset domains, validating its design goal of domain-invariant underwater enhancement.

\begin{figure*}[h]
    \centering
    \setlength{\tabcolsep}{1pt} 
    \renewcommand{\arraystretch}{0} 
    \begin{tabular}{cccccc}
        \includegraphics[width=0.16\textwidth,height=2.35cm]{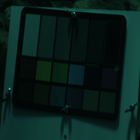} &
        \includegraphics[width=0.16\textwidth,height=2.35cm]{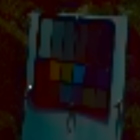} &
        \includegraphics[width=0.16\textwidth,height=2.35cm]{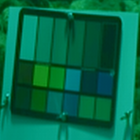} &
        \includegraphics[width=0.16\textwidth,height=2.35cm]{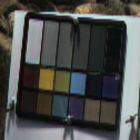} &
        \includegraphics[width=0.16\textwidth,height=2.35cm]{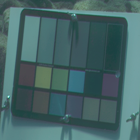} &
        \includegraphics[width=0.16\textwidth,height=2.35cm]{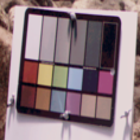} \\ [5pt]
        \small{(a) Raw Image} &
        \small{(b) P2CNet} &
        \small{(c) Phaseformer} &
        \small{(d) WaterNet} &
        \small{(e) UDNet} &
        \small{(f) DIVER (Ours)} \\
    \end{tabular}
    \caption{\small Comparison of restored color charts across different underwater enhancement methods.}
    \label{fig:both_palatte}
\end{figure*}

\begin{table*}[h]
\caption{Comparison of GPMAE for the SeaThru dataset sample images, here lower angular error values indicate a better performance.  We represent the first best scores in \textcolor{blue}{BLUE} and second best score in \textcolor{orange}{ORANGE}.}
\label{GPMAE scores}
\renewcommand{\arraystretch}{1.25}
\resizebox{\textwidth}{!}{%
\begin{tabular}{ccccccccccccc}
\hline
\textbf{Image} &
  \textbf{Raw} &
  \textbf{Sea-Thru} &
  \textbf{IBLA} &
  \textbf{DCP} &
  \textbf{UDCP} &
  \textbf{ULAP} &
  \textbf{UST} &
  \textbf{P2CNet} &
  \textbf{Phaseformer} &
  \textbf{Waternet} &
  \textbf{UDnet} &
  \textbf{DIVER} \\ \hline
\textbf{D1\_3272} &
  26 &
  8 &
  19 &
  42 &
  36 &
  35 &
  \textbf{\textcolor{blue}{4}} &
  17 &
  36 &
  \textbf{\textcolor{orange}{6}} &
  15 &
  7 \\ \hline
\textbf{D2\_3647} &
  26 &
  \textbf{\textcolor{orange}{8}} &
  24 &
  36 &
  33 &
  44 &
  13 &
  26 &
  36 &
  11 &
  15 &
  \textbf{\textcolor{blue}{7}} \\\hline
\textbf{D3\_4910} &
  22 &
  8 &
  12 &
  37 &
  30 &
  34 &
  5 &
  8 &
  35 &
  \textbf{\textcolor{orange}{5}} &
  12 &
  \textbf{\textcolor{blue}{3}} \\\hline
\textbf{D4\_0209} &
  23 &
  5 &
  17 &
  35 &
  30 &
  34 &
  \textbf{\textcolor{orange}{4}} &
  21 &
  35 &
  6 &
  18 &
  \textbf{\textcolor{blue}{2}} \\ \hline
\textbf{D5\_3374} &
  \begin{tabular}[c]{@{}c@{}}17/16\\ /15/17\end{tabular} &
  \begin{tabular}[c]{@{}c@{}}\textbf{\textcolor{orange}{4}}/3\\ /5/3\end{tabular} &
  \begin{tabular}[c]{@{}c@{}}13/19\\ /22/19\end{tabular} &
  \begin{tabular}[c]{@{}c@{}}36/36\\ /36/36\end{tabular} &
  \begin{tabular}[c]{@{}c@{}}31/36\\ /36/42\end{tabular} &
  \begin{tabular}[c]{@{}c@{}}55/55\\ /55/55\end{tabular} &
  \begin{tabular}[c]{@{}c@{}}5/\textbf{\textcolor{orange}{3}}\\  \textcolor{orange}{5/3}\end{tabular} &
  \begin{tabular}[c]{@{}c@{}}34/35\\ /36/37\end{tabular} &
  \begin{tabular}[c]{@{}c@{}}36/36\\ /36/36\end{tabular} &
  \begin{tabular}[c]{@{}c@{}}9/12\\ /12/13\end{tabular} &
  \begin{tabular}[c]{@{}c@{}}13/15\\ /15/15\end{tabular} &
  \begin{tabular}[c]{@{}c@{}}\textcolor{blue}{1/3}\\\textcolor{blue}{/4/2} \end{tabular} \\ \hline
\end{tabular} 
}
\end{table*}

\subsection{Quantitative Evaluation}
The quantitative comparisons of the DIVER and SOTA methods are presented in Tables ~\ref{tab:Non-ref scores_1} - \ref{GPMAE scores} across diverse evaluation settings. We evaluate our method on two categories of benchmarks: unpaired datasets (SeaThru, OceanDark, USOD10K, FishTrac, and U45), where ground-truth reference images are unavailable, and paired datasets (UIEB, LSUI, and EUVP), where each raw underwater image is accompanied by a corresponding clean target. 
 On the unpaired benchmark datasets presented in Table~\ref{tab:Non-ref scores_1}, DIVER achieves consistently better UIQM, UCIQE, and BRISQUE scores on the SeaThru, OceanDark, USOD10K, FISHTRAC and U45 datasets. It surpasses both traditional and deep learning approaches on OceanDark, USOD10K, FISHTRAC, and U45 in terms of UCIQE, while attaining comparable performance in UIQM. Specifically, considering UCIQE metric, DIVER shows at least 26\% improvement over traditional methods (IBLA, DCP, UDCP, ULAP, RGHS), 9.6\% over methods using transformer architecture like P2CNet and Phaseformer, 25\% over UST, 9\% over WaterNet , and 32\% over UDNet. On the UIQM metric, DIVER records 18\% and 23\% improvements on FISHTRAC when compared to P2CNet and WaterNet respectively, and 29\% and 11\% improvements on USODD10K relative to Phaseformer and UDNet respectively. While BRISQUE scores show mixed results, DIVER maintains competitive performance with best scores on SeaThru and U45 and second best on OceanDark dataset. From Table~\ref{tab:Non-ref scores_1}, it can be observed that some methods perform slightly better than DIVER in terms of UIQM scores, particularly UDNet which achieves higher values on U45. However, higher quantitative scores do not necessarily correlate with better perceptual quality \cite{li2019}, and DIVER frequently produces visually good results illustrated in Figures \ref{fig:comparison_all_1}-\ref{fig:comparison_all_2}, even when numerical differences are marginal.
 

\begin{figure}
    \centering 
    \setlength{\tabcolsep}{1pt}
    \begin{tabular}{cc}
        \textbf{(a)~\textit{Raw Image}} & \textbf{(b)~\textit{Enhanced Image}} \\

        \includegraphics[width=0.45\linewidth,height=0.85in]{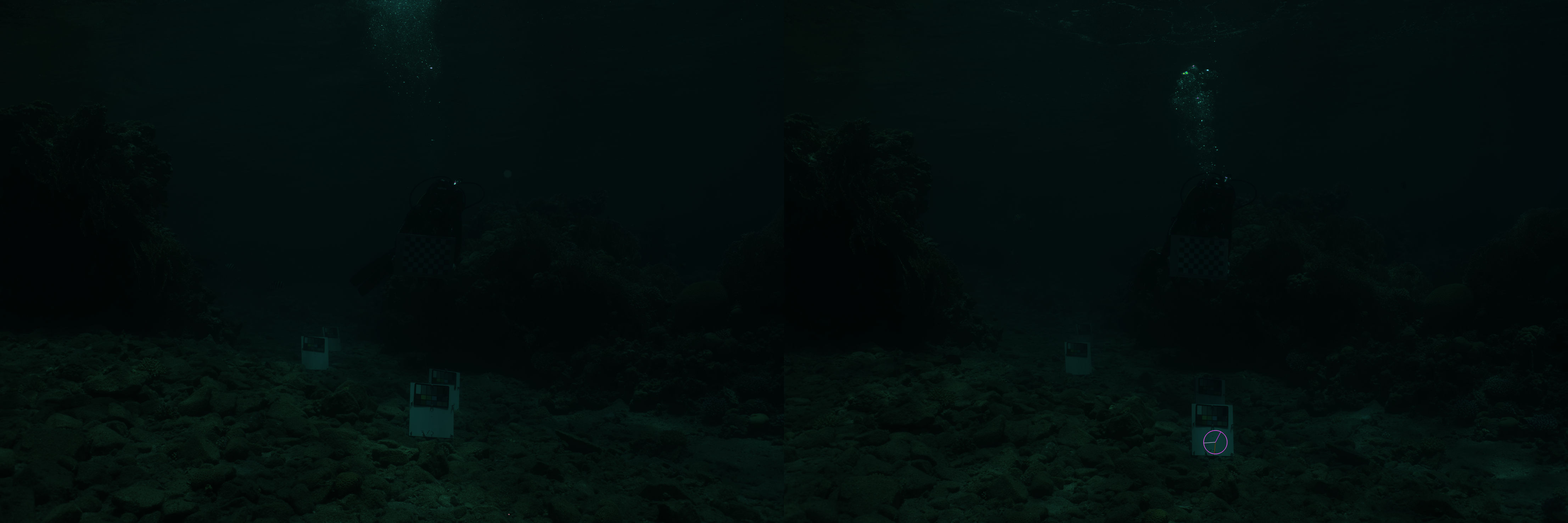} &
        \includegraphics[width=0.45\linewidth,height=0.85in]{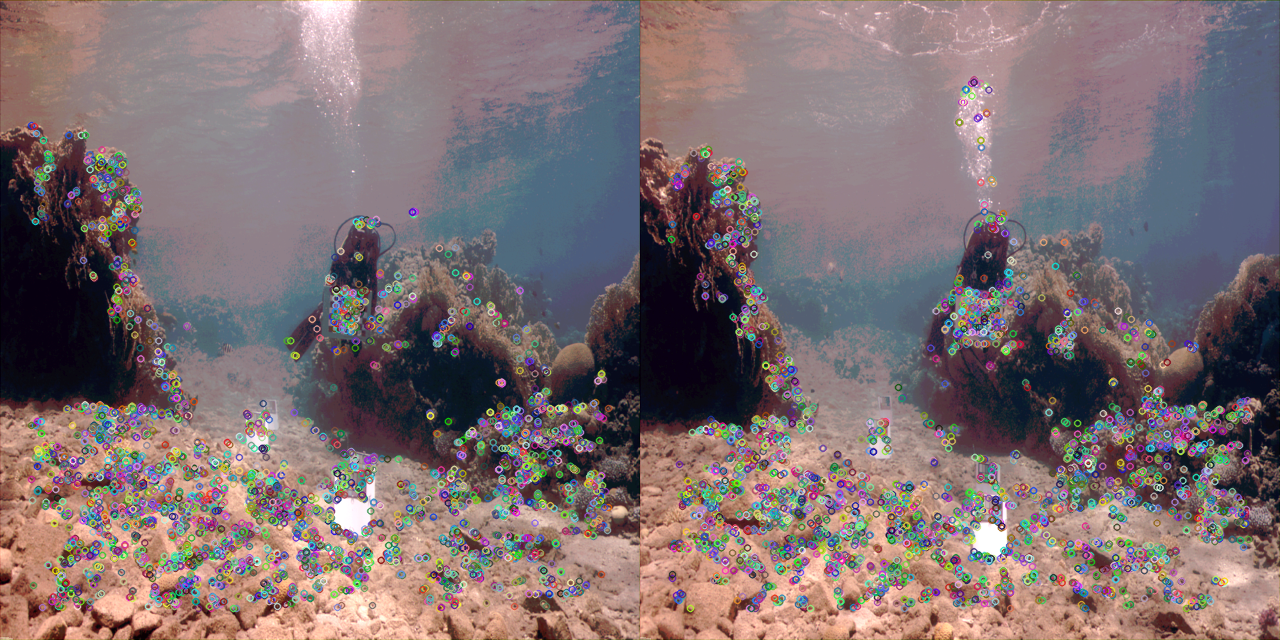} \\ [-1.5ex]

        \includegraphics[width=0.45\linewidth,height=0.85in]{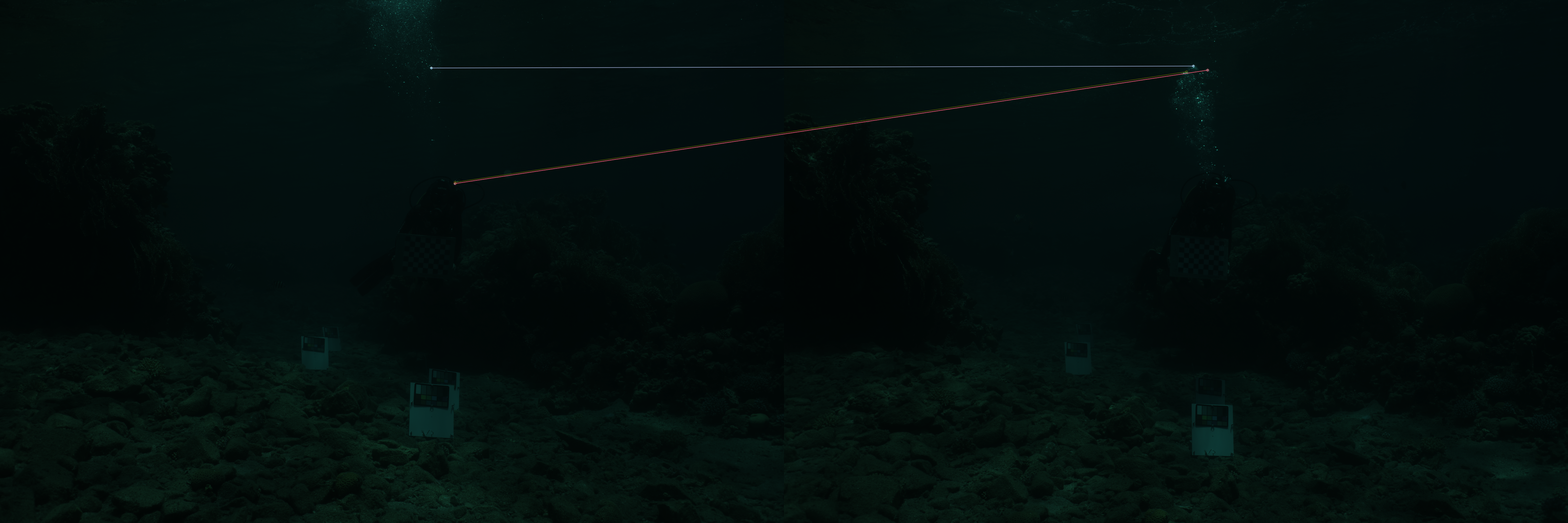} &
        \includegraphics[width=0.45\linewidth,height=0.85in]{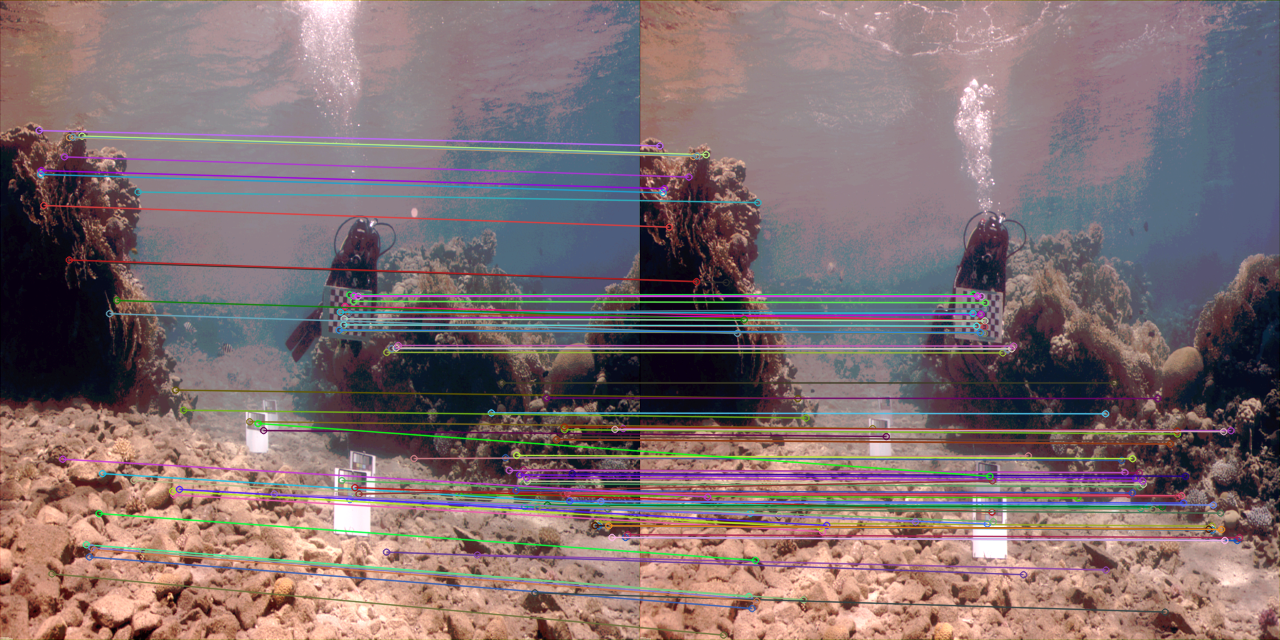} \\

        \includegraphics[width=0.45\linewidth,height=0.85in]{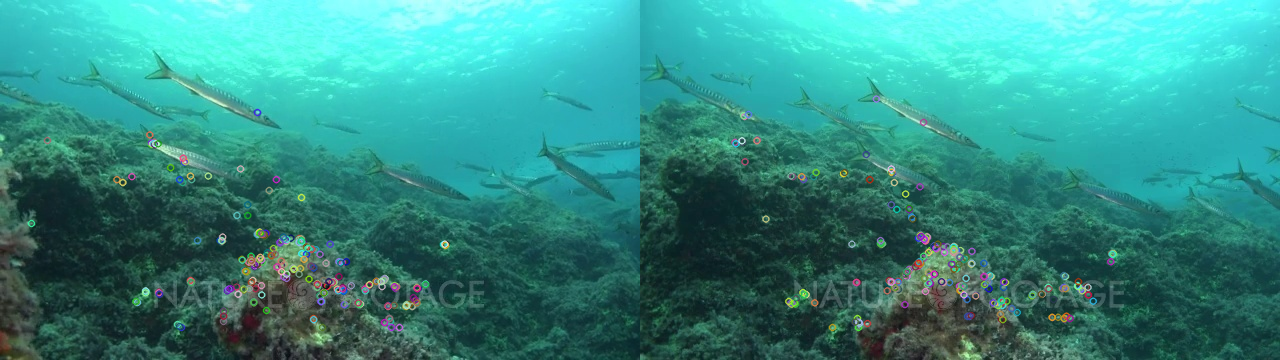} &
        \includegraphics[width=0.45\linewidth,height=0.85in]{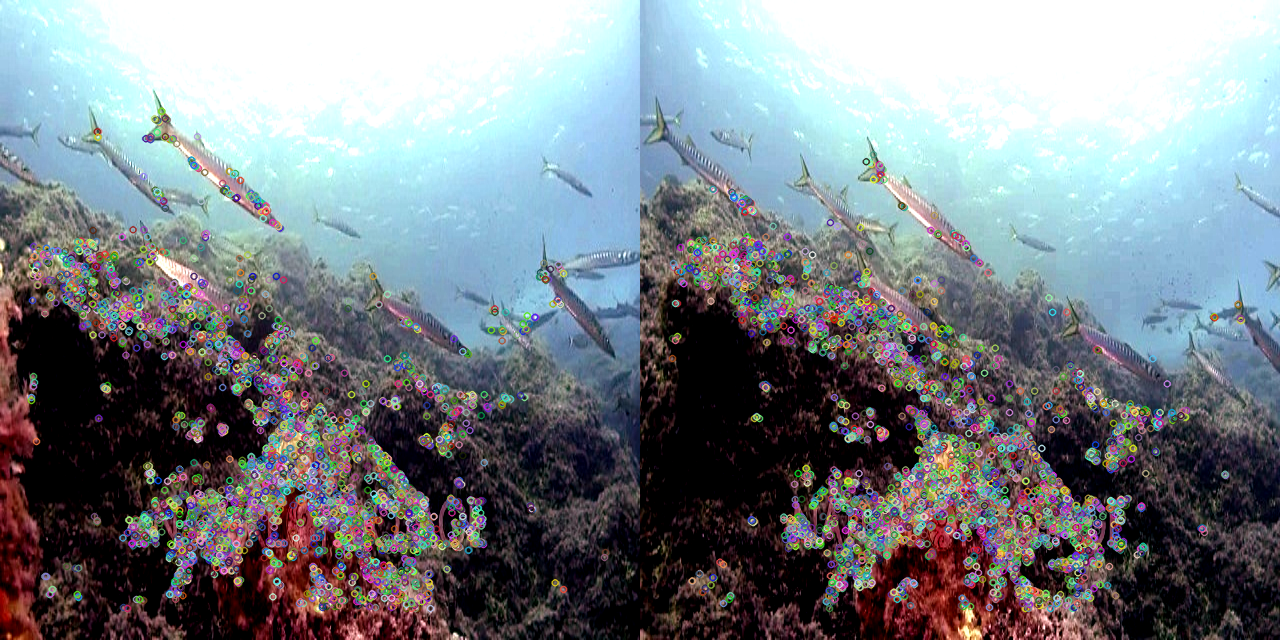} \\ [-1.5ex]

        \includegraphics[width=0.45\linewidth,height=0.85in]{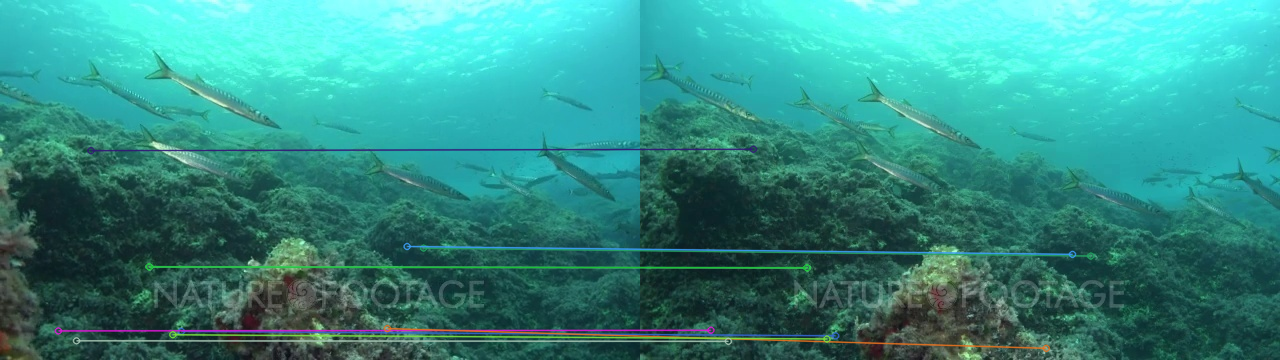} &
        \includegraphics[width=0.45\linewidth,height=0.85in]{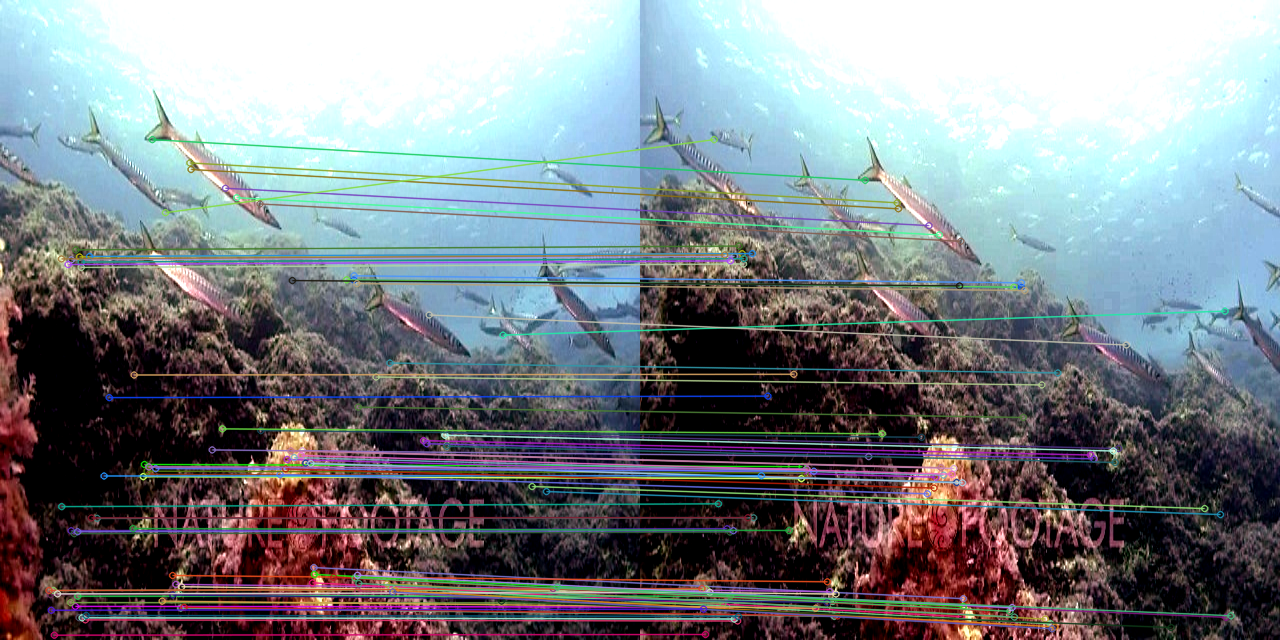} \\
    \end{tabular}
    \caption{Comparison of key points and image feature matching for SeaThru (row 1 and 2) and LSUI (row 3 and 4) datasets between (a) Raw Image and (b) Enhanced Image with  DIVER method.}
    \label{sift}
\end{figure}
 For the paired dataset summarized in Table ~\ref{Ref scores_1}, DIVER shows competitive results on the UIEB, LSUI, EUVP, and UFO-120 datasets. Notably, it achieves the highest PSNR value on UIEB and second highest value on LSUI, EUVP and UFO-120 datasets, the highest SSIM on UIEB and EUVP, the best UCIQE scores on UIEB, LSUI and EUVP and comparable UIQM scores with second highest scores on EUVP and UFO-120. For PSNR, the most substantial gains are observed against conventional methods like UDCP with 54\% on UIEB, 58\% on LSUI, and 50\% on EUVP, and against ULAP with 29\% on UIEB and 57\% on LSUI. The smallest improvements are against WaterNet on UIEB with around 3\% improvement. In terms of UCIQE scores, DIVER demonstrates consistent improvement of 0.3\% to 59\%, where it outperforms almost all methods across all four paired datasets. Comparing SSIM and UIQM scores, DIVER exhibits mixed results on UIEB and EUVP, DIVER achieves the best SSIM with at least 2\% improvement over competing methods, but on LSUI and UFO-120, methods like WaterNet and Phaseformer outperform DIVER. For UIQM, DIVER shows strong performance with at least 13\% improvements over conventional methods on UIEB, but shows comparable results against deep learning techniques with maximum of 25\% improvement. Nevertheless, on datasets like UFO-120 and LSUI, DIVER remains competitive, and irrespective of scores, DIVER produces consistent visually dominant results as demonstrated in Figures \ref{fig:comparison_all_2} and \ref{fig:comparison_all_3}. Although DIVER was developed as an unsupervised method, it outperforms several supervised deep learning based models trained on these specific datasets, particularly excelling in PSNR and UCIQE metrics. These result suggests that DIVER is better at preserving structural information, color fidelity, and contrast than supervised methods such as WaterNet.


The GPMAE results in Table~\ref{GPMAE scores} further validate the better color restoration capability of DIVER on the SeaThru dataset. Since GPMAE directly measures chromatic deviation from the neutral-gray target, this evaluation serves as a reliable indicator of true underwater color recovery. This a critical validation step where no ground-truth clean images exist.  Across all five scenes from SeaThru dataset, DIVER consistently achieves the lowest angular error, reflecting more accurate neutral-gray color recovery compared to both conventional and deep learning baselines. On average, DIVER reduces angular error by 74\% versus IBLA, 84\% versus DCP, and 87\% versus ULAP, demonstrating substantial improvement over physics-based methods. Against learning-based competitors, DIVER achieves 30\% lower error than WaterNet, 68\% lower than UDNet and 17\% lower error than UST. This highlights the robustness of DIVER even in challenging low-light conditions. Further, DIVER surpasses even Sea-Thru (ST) method, a physics-calibrated reference algorithm, with 37\% lower angular deviation. Because GPMAE directly quantifies how closely colors align to their true neutral reference, it is a critical indicator of genuine color restoration capability in underwater enhancement. These improvements are visually reinforced in Figure~\ref{fig:both_palatte}, where DIVER uniquely restores a stable and realistic color palette without the green/red dominance. Deep learning methods like Phaseformer show noticeable green channel dominance, while UDNet produces hazy palatte. WaterNet shows more saturation making the palette daraker. In contrast, DIVER restores a reference-like chromatic distribution across all palette patches. 


Beyond quantitative and quantitative analysis, it is equally important to evaluate how enhancement influences real robotic perception tasks. Therefore, the next section investigates the impact of DIVER on feature detection and matching performance for underwater autonomy.

\subsection{Effect of Enhancement on Robot Perception}

A key objective of underwater image enhancement is to strengthen the visual perception capabilities of underwater robots, which is essential for reliable autonomous decision-making in complex environments. To quantitatively evaluate DIVER’s contribution to perception, we employ feature detection and keypoint matching—core components in tasks such as robotic navigation, object detection, and 3D reconstruction. Specifically, we utilize the Oriented FAST and Rotated BRIEF (ORB) \cite{Rublee2011ORB} to extract local keypoints and apply the Random Sample Consensus (RANSAC) algorithm to establish robust correspondences between image pairs. As illustrated in Fig.~\ref{sift}, the enhanced images exhibit a substantial increase in both detected features and successful matches compared to their raw counterparts. For instance, in the ST dataset, the enhanced images achieved 1,436 matched keypoints versus only 8 in the raw input, while the LSUI dataset showed an increase from 94 to 1,625 matches. This simultaneous rise in total keypoints and reliable correspondences indicates that DIVER significantly improves the visibility of structural details.

\begin{table}[h]
\caption{Ablation study: score comparison against different blocks of DIVER on SeaThru dataset.}
\label{TAB:Non-ref_abalation}
\normalsize
\begin{tabular}{lll}

\hline
\multicolumn{1}{c}{\textbf{Blocks Proposed}} & \multicolumn{1}{c}{\textbf{UCIQE}} & \multicolumn{1}{c}{\textbf{UIQM}}  \\ \hline
\textbf{Raw}                                 &    0.1062                                &  	0.9980                                                                       \\ \hline
\textbf{IlluminateNet}                       &    0.7007                                &     2.3685                                                                    \\ \hline
\textbf{IlluminateNet + AOCM}                                &  0.5783                                &    2.5879                                                                  \\ \hline
\textbf{IlluminateNet + AOCM + HON}                                 &      \textbf{0.8470}                               &    \textbf{2.8685}                                                                   \\ \hline

\end{tabular}
    
\end{table}
\begin{table}[h]
\caption{Ablation study: score comparison against different blocks of DIVER on UFO-120 dataset.}
\label{TAB:ref_abalation}
\normalsize
\begin{tabular}{llll}
\hline
\multicolumn{1}{c}{\textbf{Blocks Proposed}} & \multicolumn{1}{c}{\textbf{PSNR}} & \multicolumn{1}{c}{\textbf{SSIM}} & \multicolumn{1}{c}{\textbf{UCIQE}} \\ \hline
\textbf{Raw}                                 & 12.67      & 0.5365                                                                    & 0.9160                                   \\ \hline
\textbf{SEF}                 &13.66      &0.5384                                              & 0.9308                                   \\ \hline
\textbf{SEF + AOCM}             &21.70    &  0.6051                                                     & 0.7430                                   \\ \hline
\textbf{SEF + AOCM + HON}                &\textbf{23.69 }   &  \textbf{0.6079 }                                                                                       & \textbf{0.9620}                                   \\ \hline
\end{tabular}
\end{table}

While the perception experiments validate the practical impact of DIVER in real-world robotic scenarios, it is equally important to understand how each architectural component contributes to this performance. The subsequent ablation study examines the role of each enhancement module, demonstrating how the progressive pipeline contributes to improved restoration quality.

\begin{figure*}[h]
    \centering
    \setlength{\tabcolsep}{0pt} 
    \renewcommand{\arraystretch}{0} 

    \begin{tabular}{cccc@{\hspace{0.5cm}}cccc} 
        \includegraphics[width=0.12\textwidth,height=2.35cm]{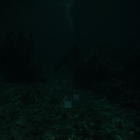} &
        \includegraphics[width=0.13\textwidth,height=2.35cm]{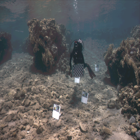} &
        \includegraphics[width=0.12\textwidth,height=2.35cm]{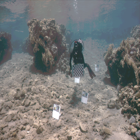} &
        \includegraphics[width=0.12\textwidth,height=2.35cm]{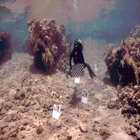} &
        \includegraphics[width=0.12\textwidth,height=2.35cm]{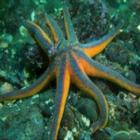} &
        \includegraphics[width=0.12\textwidth,height=2.35cm]{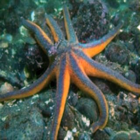} &
        \includegraphics[width=0.12\textwidth,height=2.35cm]{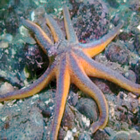} &
        \includegraphics[width=0.12\textwidth,height=2.35cm]{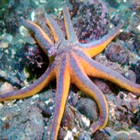} \\ [5pt]

        \small{(a)~Raw Image} & \small{(b)~IlluminateNet} & \small{(c)~AOCM} & \small{(d)~HON} &
        \small{(e)~Raw Image} & \small{(f)~SEF} & \small{(g)~AOCM} & \small{(h)~HON} \\
    \end{tabular}

    \caption{\small  Ablation study of the proposed DIVER framework across two datasets, illustrating its component blocks: (a)–(d) low‑lit images and (e)–(h) adequately lit images.}
    \label{fig:abalation_combined}
\end{figure*}


\subsection{Ablation Study}

To understand how each component of the DIVER architecture contributes to domain-invariant enhancement, we perform a structured ablation analysis under two illumination conditions: \textit{low-lit} and \textit{adequately lit} scenes. This mirrors the initial illumination assessment stage used in DIVER. For each condition, we report quantitative metrics (UCIQE, UIQM) for unpaired datasets and PSNR/SSIM for paired datasets, with results summarized in Tables~\ref{TAB:Non-ref_abalation} and~\ref{TAB:ref_abalation}. Representative qualitative comparisons is illustrated in Figure~\ref{fig:abalation_combined}.

\paragraph{Low-Lit Ablation (IlluminateNet branch).}
As shown in Table~\ref{TAB:Non-ref_abalation} and depicted in (a) - (d) blocks of Figure~\ref{fig:abalation_combined}, applying IlluminateNet alone dramatically improves UCIQE (from $0.1062$ to $0.7007$) and UIQM (from $0.9980$ to $2.3685$), demonstrating its effectiveness in restoring luminance and spectral balance. Adding AOCM further enhances UIQM (from $2.3685$ to $2.5879$), indicating improved contrast and color uniformity, although UCIQE temporarily decreases due to the stronger luminance normalization. With Hydro-OpticNet (HON), we achieve the best scores (UCIQE $0.8470$, UIQM $2.8685$), confirming that physics-guided backscatter and attenuation correction are essential for restoring natural color and global image quality .

\paragraph{Adequately Lit Ablation (SEF branch).}
For well-lit images (Table~\ref{TAB:ref_abalation}), SEF alone yields modest gains in PSNR and UCIQE, primarily correcting channel dominance. Introducing AOCM produces a significant change in PSNR (from $13.66$ to $21.70$ dB), reflecting strong improvements in contrast and structural clarity observed in blocks (e) - (h) of Figure~\ref{fig:abalation_combined} . However UCIQE decreases due to aggressive contrast flattening. Once Hydro-OpticNet is included, PSNR, SSIM, and UCIQE all reach their highest values, confirming that the physics-driven correction resolves residual wavelength attenuation and scattering that empirical modules cannot fully address.

Across both branches, AOCM consistently improves contrast and perceptual sharpness, while Hydro-OpticNet provides the final radiometric refinement necessary for domain-invariant performance. The complementary behavior of empirical and physics-guided modules highlights the necessity of DIVER's hierarchical design.

In summary, the experimental evaluation demonstrates that DIVER shows consistently better enhancement quality across diverse underwater environments when compared to a broad set of traditional and deep learning baselines. Qualitative comparisons highlight natural color fidelity, haze removal, and structural clarity across variable lighting and water types, while quantitative results confirm best and near best performance on both paired and unpaired datasets. Robot perception tests further show that the enhanced imagery significantly improves feature detection and matching reliability, supporting downstream autonomy tasks. The ablation study verifies that each component in the DIVER architecture plays a distinct and complementary role, with the physics-guided modules proving crucial for radiometric generalization. Together, these findings validate DIVER as a robust and domain-invariant solution for underwater image enhancement.

\section{Conclusion}\label{s5}
In this paper, we proposed DIVER, an unsupervised and domain-invariant underwater image restoration architecture designed to overcome the optical degradation caused by absorption, scattering, and illumination non-uniformity across diverse underwater environments. DIVER integrates four complementary components: IlluminateNet for depth-guided luminance and spectral recovery in low-light scenes, the Spectral Equalization Filter (SEF) for channel balancing in adequately lit images, the Adaptive Optical Correction Module (AOCM) for contrast enhancement and speckle suppression, and Hydro-OpticNet (VeilNet and AttenNet) for physics-constrained compensation of backscatter and wavelength-dependent attenuation.

Comprehensive experiments on nine public datasets demonstrate that DIVER consistently improves restoration quality over classical and deep learning baselines. Across unpaired datasets, DIVER achieves 22–87\% UCIQE gains over traditional priors and 24–74\% improvements relative to transformer-based architectures, while performing 34–50\% better than UDNet. On paired datasets, it attains the highest PSNR and SSIM scores on UIEB and top UCIQE scores across UIEB, LSUI, EUVP, and UFO-120, despite operating in a fully unsupervised setting. While UCIQE and UIQM measure perceptual quality, GPMAE uniquely quantifies fidelity of color restoration by evaluating angular error of neutral (gray) patches. GPMAE results on SeaThru further confirm enhanced color fidelity with 4.9–98\% lower angular errors than competing methods. In addition to numeric gains, qualitative evaluations show improved haze removal, balanced color correction, and sharper structure recovery. 

Ablation results further validate the role of each component: IlluminateNet/SEF provide luminance and spectral balancing improvements, AOCM refines contrast and chromatic stability, and Hydro-OpticNet delivers the largest gain by modeling depth-dependent scattering and attenuation yielding up to 87\% UCIQE improvement (SeaThru dataset) and 79\% PSNR gain (UFO-120 dataset) over the raw input. Additionally, enhanced feature-matching in our perception tests highlights its practical utility for AUV/ROV visual navigation. Together, these results establish DIVER as a generalizable and radiometrically consistent solution for underwater visual enhancement in real-world deployment.
As DIVER focuses primarily on depth-dependent attenuation correction, its effectiveness can be reduced in regions with limited depth cues, such as distant backgrounds. In future, we will be exploring improved background restoration techniques by integrating scene-aware priors to recover large-range structures. 


\end{document}